\documentclass[10pt,conference,letterpaper]{IEEEtran}

\let\labelindent\relax
\usepackage{array,multirow,times}
\usepackage{subfigure}
\usepackage{caption}
\usepackage{amsmath}
\usepackage{url} 

\makeatletter
\newcommand{\thickhline}{%
    \noalign {\ifnum 0=`}\fi \hrule height 1pt
    \futurelet \reserved@a \@xhline
}

\newcolumntype{"}{@{\hskip\tabcolsep\vrule width 1pt\hskip\tabcolsep}}
\makeatother

\usepackage{pifont}

\usepackage[ruled,vlined,linesnumbered]{algorithm2e}
\usepackage{setspace}

\SetCommentSty{mycommfont}
\usepackage{tikz}

\usepackage{array,multirow}
\usepackage{caption}

\usepackage{etoolbox}

 \makeatletter
 \preto{\@verbatim}{\topsep=0pt \partopsep=0pt }

\usepackage{flushend}
\usepackage[numbers,sort]{natbib}
 \usepackage{enumitem}

\newtheorem{defn}{Definition}

\newtheorem{example}{Example}

\usepackage{cleveref}
\usepackage{mathtools}
\DeclarePairedDelimiter\ceil{\lceil}{\rceil}
\DeclarePairedDelimiter\floor{\lfloor}{\rfloor}

\title{FAST: Frequency-Aware Spatio-Textual Indexing for In-Memory Continuous Filter Query Processing}
\author{%
{Ahmed R. Mahmood{\small $~^{\#}$}, Ahmed M. Aly{\small $~^{*}$}, Walid G. Aref{\small $~^{\#}$}}%
\vspace{1.6mm}\\
\fontsize{10}{10}\selectfont\itshape
\hspace{4mm}$^{\#}$\,Purdue University, West Lafayette, IN  \hspace{30mm}$^{*}$\,Google Inc., Mountain View, CA \\
\fontsize{9}{9}\selectfont\ttfamily\upshape
$^{\#}$\,\{amahmoo,aref\}@cs.purdue.edu  \hspace{42mm}$^{*}$\, aaly@google.com%
\vspace{1.2mm}\\
\fontsize{10}{10}\selectfont\rmfamily\itshape
}
\begin{document}
\maketitle
\begin{abstract} 
\sloppy
Many applications need to process massive streams of spatio-textual data in real-time against continuous spatio-textual queries. For example, in location-aware ad targeting publish/subscribe systems, it is required to disseminate millions of ads and promotions to millions of users based on the locations and textual profiles of users. In this paper, we study indexing of continuous spatio-textual queries. 
There exist several related spatio-textual indexes that typically integrate a spatial index
with a textual index.
However, these indexes usually have a high demand for main-memory and assume that the entire vocabulary of keywords is known in advance.
Also, these indexes do not successfully capture the variations 
in
the frequencies of keywords across different spatial regions and treat frequent and infrequent keywords in the same way. 
Moreover, existing indexes do not adapt to the changes in workload over space and time. For example, some keywords may be trending at certain times in certain locations and this may change as time passes. This affects the indexing and searching performance of existing indexes significantly. 
In this paper, we introduce FAST, a {\em F}requency-{\em A}ware {\em S}patio-{\em T}extual index for continuous spatio-textual queries. FAST is a main-memory index that requires up to one third of the memory needed by the state-of-the-art index.
FAST does not assume prior knowledge of the entire vocabulary of indexed objects. FAST adaptively accounts for the difference in the frequencies of keywords within their corresponding spatial regions to automatically choose the best indexing approach that optimizes the insertion and search times. Extensive experimental evaluation using real and synthetic datasets demonstrates that FAST is up to 3x faster in search time and 5x faster in insertion time than the state-of-the-art indexes.
\end{abstract}
\sloppy
\section{Introduction}
Nowadays, many applications rely on processing and analyzing spatio-textual data. 
Example applications include social networks (e.g., Facebook), micro-blogs (e.g., Twitter), web search for local places and events, and location-aware ad targeting. These applications process spatio-textual data at a massive scale and in real-time. For example, 500 million tweets~\cite{wang2015ap,geotaggedtweets} and 9 million Foursquare check-ins~\cite{foursquare} are being generated and processed daily. 
These applications require efficient spatio-textual indexing to support 
this
scale of spatio-textual data.

In this paper, we 
focus on the indexing of continuous spatio-textual filter queries. This type of queries appears in many applications, e.g., location-aware publish/subscribe systems~\cite{mahmood2015tornado}, information dissemination~\cite{yan1994index}, and sponsored search~\cite{konig2009data}. A continuous spatial-keyword filter query consists of a spatial range and an associated set of keywords. For a stream of spatio-textual objects, a continuous spatio-textual filter query identifies the objects that fall inside the spatial range of the query and that contain all the keywords of the query.

\begin{figure}[t!]
 	\centering	\includegraphics[width=2.7in]{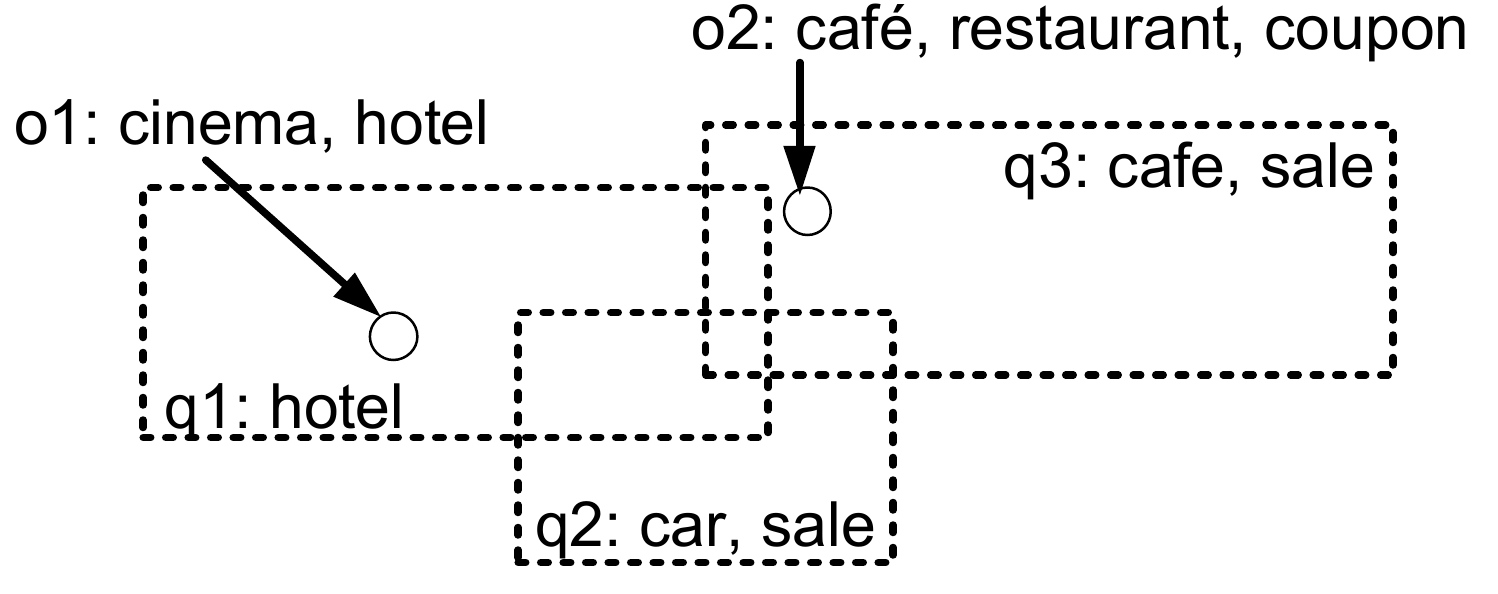}	\caption{E-coupon example.}	\label{fig:simpleexample}
\end{figure}

\begin{example} Figure~\ref{fig:simpleexample} illustrates a sample location-aware e-coupon application in a location-aware publish/subscribe system. Three users show interest in promotions represented by the three continuous spatio-textual queries $q1$, $q2$, and $q3$. Promotion $o1$ matches Query $q1$ because $o1$ is located inside $q1$'s spatial range,    and contains all the keywords of $q1$.
\end{example}

\begin{figure}[t!]
	\centering	\includegraphics[width=1.7in]{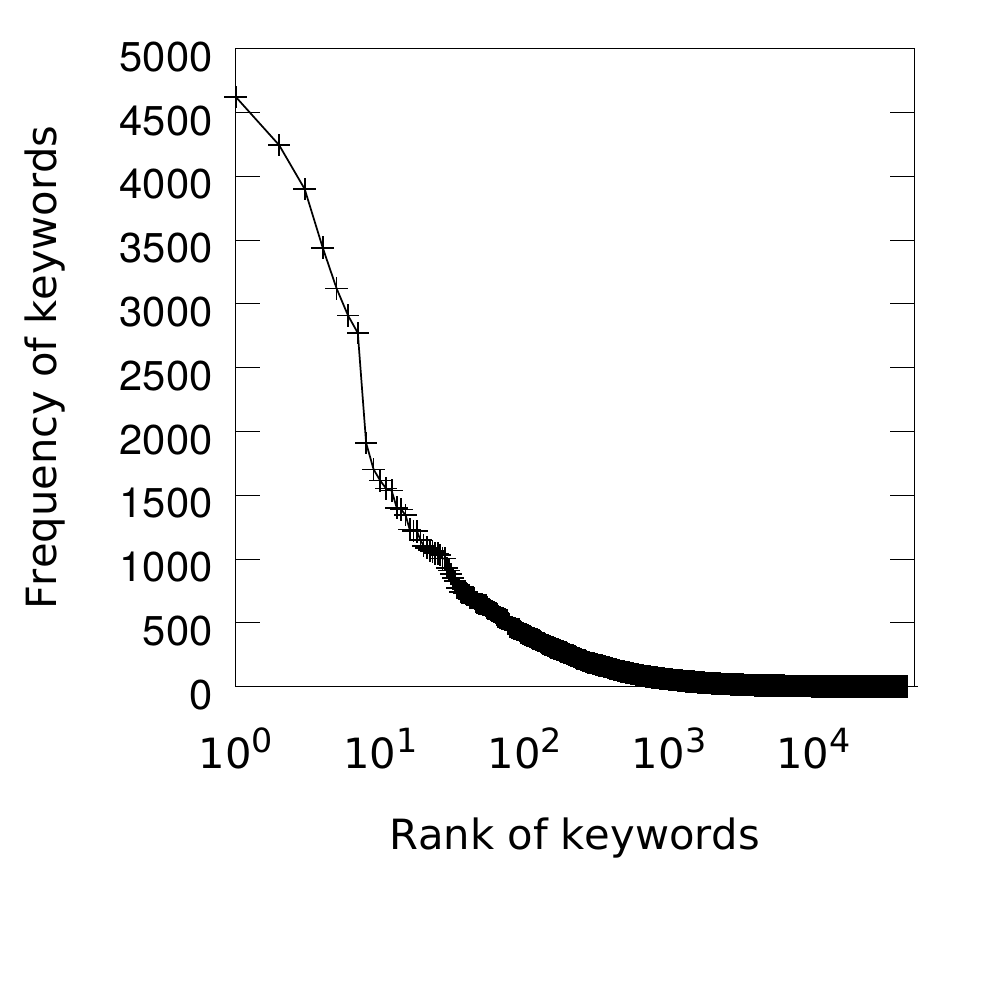}	\caption{Zipfian distribution of query keywords.}	\label{fig:zipf}
\end{figure}

\begin{figure}[t!]
	\centering	\includegraphics[width=3in]{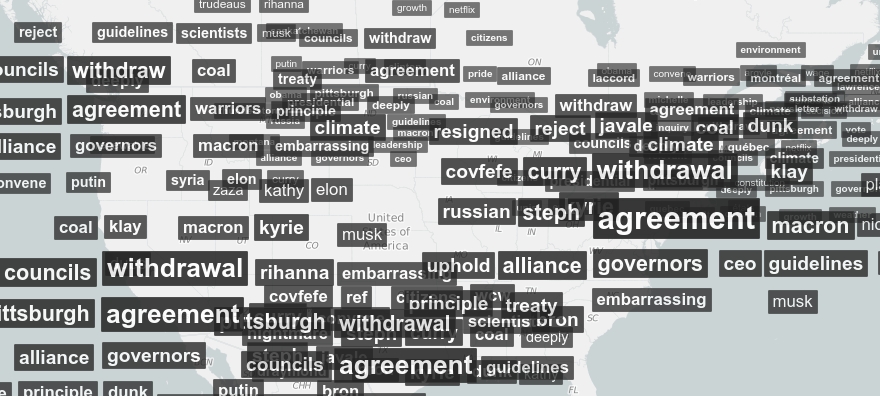}	\caption{Spatial distribution of popular keywords in tweets\protect\footnotemark within the United States.}	\label{fig:spatialtrends}
\end{figure}

Recently, several access-methods have been proposed to handle continuous spatio-textual queries in streaming environments, e.g.,~\cite{wang2015ap,chen2013efficient,li2013location}. These access methods integrate a spatial index (e.g., a spatial grid,  the R-tree~\cite{guttman1984r}, or the quad-tree~\cite{finkel1974quad}) with a textual index (e.g., the  inverted list~\cite{zobel2006inverted}, or the ordered-keyword trie~\cite{hmedeh2012subscription}). However, these access-methods do not account for the frequencies and the popularity of some of the keywords within the indexed spatio-textual queries. Consider Figure~\ref{fig:zipf} that illustrates the frequencies of keywords in a set of 50,000 tweets. 
The frequencies of the keywords follow a Zipfian distribution~\cite{powers1998applications}. This distribution has many infrequent keywords and few frequent keywords. Although the distribution of the frequencies of keywords is Zipfian, the exact ranking and frequencies of keywords may not be known and the frequencies of keywords may change overtime. Also, new keywords get introduced to the vocabulary and it is estimated that 1000 new words are added to the Oxford dictionary every year\footnote{http://blog.oxforddictionaries.com/august-2013-update}. Also, some infrequent keywords may become frequent, e.g., Hurricane Irma. Furthermore, the distribution of the frequent keywords is non-uniform across the space as illustrated in Figure~\ref{fig:spatialtrends}.


Existing indexes treat queries with 
frequent
keywords 
in 
the same way as 
it treats 
queries 
that contain
infrequent keywords. For example, when using inverted lists~\cite{zobel2006inverted}, a query is indexed based on a single keyword. This keyword is usually the least-frequent keyword. Inverted lists are well-suited for queries with infrequent keywords. However, the inverted list structure has the following two limitations: (1)~it suffers from poor performance for queries that only have frequent keywords 
because the inverted 
lists associated with these frequent keywords can be very large, and (2)~it assumes the knowledge of the entire vocabulary of keywords and their frequencies. However, in real scenarios, e.g., when processing tweets, the entire vocabulary and the ranking of the keywords 
are not known a priori. 

Another popular textual index is the ordered-keyword 
trie~\cite{hmedeh2012subscription} that is a variation of the traditional trie structure~\cite{knuth1968art}. The ordered-keyword trie indexes keywords instead of characters in the traditional trie structure. The ordered-keyword trie offers better textual filtering for queries with no infrequent keywords. However, the ordered-keyword trie 
suffers from the following limitations: (1)~it has 
a large memory-footprint, and (2)~it does not quickly prune queries with infrequent keywords unless the indexed keywords have a total order based on their frequencies. Having a total order of keywords based on their frequencies requires prior knowledge of the entire vocabulary of keywords and their frequencies, which may not be feasible.  
It is challenging to support efficient indexing of continuous spatio-textual queries in a streaming environment due to the following reasons:
\begin{itemize}
\item The massive scale of the indexed queries as it is typical to deal with millions of rapidly arriving continuous queries. 
\item Spatio-textual objects are streamed at a high rate, and it is required to process these objects against millions of indexed queries with minimal latency.
\item The locations and frequencies 
of spatio-textual data and queries are not uniformly distributed. Hence, an efficient index needs to account for the varying distributions of spatial and textual aspects of 
the
indexed queries.
\item The assumption of knowing the entire vocabulary of keywords in advance is not valid in many situations, e.g., as 
in
processing social media posts. 
\end{itemize}

\footnotetext{https://www.trendsmap.com}

To address these challenges, we introduce FAST, a {\em F}requency-{\em  A}ware {\em S}patio-{\em T}extual access method for indexing continuous queries in a streaming environment. 
FAST is designed as a main-memory index to minimize indexing and searching time and to meet the real-time processing requirements of rapidly-arriving spatio-textual data and queries. 

FAST 
treats
the frequencies of keywords and their distribution in space as first-class properties of spatio-textual queries. 
FAST integrates a 
variant 
of the incomplete \textit{spatial pyramid} structure~\cite{aref1990efficient} with a 
new
textual index,
termed 
\textit{the adaptive keyword index ($AKI$)} to boost the spatial and textual pruning power of FAST. The spatial pyramid is a multi-resolution spatial index that is being adopted in many spatio-textual indexes, 
e.g.,
~\cite{magdy2015towards,lee2015processing}.
AKI accounts for the frequencies of 
the
keywords, and automatically distinguishes between 
frequent and infrequent
keywords. 
AKI allows FAST to 
quickly prune queries that have infrequent keywords. Queries that have no infrequent keywords are indexed in a more selective way in FAST. 
Moreover, instead of searching for all the keywords at all the levels of the pyramid, FAST adopts  \textit{frequency-aware spatial indexing},  
where queries containing infrequent keywords are indexed only at the top level
of the spatial pyramid. This reduces the number of keywords being searched 
while descending 
the spatial pyramid. Because of this frequency awareness, FAST is 3x faster than the state-of-the-art 
indexes
in terms of search time.

 
The textual index AKI is designed to reduce the memory footprint of FAST by distinguishing between queries with no infrequent keywords and queries that have some infrequent keywords. 
FAST requires less memory for queries that have some infrequent keywords by attaching the queries 
only 
to the least-frequent keyword
and not to all keywords in the query. 
Also, FAST improves the pruning power for queries with no infrequent keywords by attaching these queries to longer sequences of 
cascaded keywords
that appear in the query.
Hence, FAST demands more space only when higher pruning power is needed.
When queries span multiple spatial nodes inside FAST's spatial pyramid, FAST adopts a spatial sharing technique to further reduce its memory footprint.
These optimizations 
results in reducing
the memory footprint of FAST by up to one-third of that of 
the 
state-of-the-art 
indexes.

FAST 
does not require
prior knowledge of the entire vocabulary of keywords 
or
their frequencies. FAST captures this information 
dynamically
as queries 
get inserted or deleted.
Also, FAST employs a lazy cleaning mechanism that removes 
the 
expired queries and updates the 
index structure
to reflect the current frequencies of the 
query keywords.

The main contributions of this paper are as follows:
\begin{itemize}
\item We introduce FAST, a 
frequency-aware spatio-textual index
for 
continuous spatio-textual filter queries in a streaming environment. 
FAST 
is equipped 
with a 
new
\textit{adaptive keyword index ($AKI$)} that adaptively accounts for the frequencies of keywords and does not require prior knowledge of the vocabulary of keywords 
or 
their frequencies.
\item 
FAST 
is designed 
as a light-weight index that uses a \textit{frequency-aware spatial pyramid} and \textit{spatial-sharing of query lists} to improve the performance of the search operation with an optimized memory footprint.
\item We propose a 
light-weight
cleaning mechanism that lazily removes the expired queries, and dynamically re-adjusts the structure of the index to account for changes in the frequencies of the keywords of 
the
indexed queries.   
\item We present a mathematical analysis that aids in tuning the parameters of FAST. 
\item We conduct an extensive performance study of FAST using real and synthetic datasets. When compared to the state-of-the-art 
indexes, results 
demonstrate 
that FAST 
is 3x faster in 
search time,
5x faster in 
insertion time, 
and requires up to one-third of the memory needed by the state-of-the-art index.
\end{itemize}

The rest of this paper proceeds as follows: Section~\ref{sec:Preliminaries} 
presents 
notations used throughout the paper and 
presents the
data structures related to FAST. The structure and the main algorithms of FAST are presented in Section~\ref{sec:structure}.  
The performance evaluation 
of FAST is 
presented in Section~\ref{sec:experimentalevaluation}. Section~\ref{sec:relatedwork} highlights the related work,
and Section~\ref{sec:conculsion} contains concluding remarks.
\section{Preliminaries}
\label{sec:Preliminaries}
\sloppy
\begin{table}[!t]
\centering
    \caption{Notations 
    used 
    throughout the paper.}   
    \label{tab:notations}
	{\renewcommand{\arraystretch}{1}%
    \begin{tabular}{|c|c|c}
    \hline
       {\bfseries  Notation } &{\bfseries Description }\\
       \thickhline
       $o$ & \parbox{5.5cm}{A spatio-textual data object}\\
       \hline
       $O$ & \parbox{5.5cm}{A stream of spatio-textual data objects}\\
       \hline
       $q$ & \parbox{5.5cm}{A spatio-textual query}\\
       \hline
       $Q$ & \parbox{5.5cm}{The set of indexed spatio-textual queries}\\
       \hline
       $o.loc$ & \parbox{5.5cm}{The geo-location of a data object}\\
       \hline        
       $t_{exp}$ & \parbox{5.5cm}{The 
       expiration time of a query}\\
       \hline        
       $q.MBR$ & \parbox{5.5cm}{The spatial range of 
       a
       query}\\
       \hline   
       $o.text(q.text)$ & \parbox{5.5cm}{The keyword of a data object (query)}\\
       \hline
        $|text|$ & \parbox{5.5cm}{The number of
        keywords
        in $text$} \\
       \hline
       $\theta$ & \parbox{5.5cm}{The frequent-keyword threshold}\\
        \hline     
        $N_p$ &\parbox{5.5cm}{ A spatial pyramid node}\\
       \hline   
        $currenttime$ &\parbox{5.5cm}{The current wall-clock time }\\
       \hline  
       $N_t$ & \parbox{5.5cm}{A textual node}\\
       \hline
        $RIL$ & \parbox{5.5cm}{Ranked-keyword inverted list}\\
       \hline
      $OKT$ & \parbox{5.5cm}{Ordered-keyword trie}\\
       \thickhline
    \end{tabular}}
\end{table}

In this section, we 
introduce
the problem definition, and 
describe the data structures relevant to FAST. Table~\ref{tab:notations} summarizes the notations used.

\subsection{Problem Definition}
A \textbf{spatio-textual data object}, say $o$, is of the form $o=\left[oid,~loc,~text\right]$, where $oid$ is the identifier of the object, $loc$ is the geo-location of the object, and $text$ is the set of keywords associated with the data object. 

A \textbf{continuous spatio-textual filter query}, say $q$, is of the form $q=\left[qid,~MBR,~text,~t_{exp}\right]$, where $qid$ is the identifier of the query, $MBR$ is the spatial range of the query represented as a minimum bounding rectangle, i.e., $[x_{min},y_{min},x_{max},y_{max}]$, and $text$ is the set of keywords associated with the query. The continuous query $q$ remains  registered in the index until Timestamp $t_{exp}$, where $t_{exp}$ is the 
expiration
timestamp of the query. 

For a streamed spatio-textual data object, say $o$, the objective is to match $o$ with all 
the continuous
queries that have their spatial and textual criteria satisfied 
by $o$'s location and textual data.
The formal definition of spatio-textual matching is as follows:
\begin{defn}
\textit{Spatio-Textual Matching}. A spatio-textual data object $o$ matches a continuous spatio-textual query $q$ when the spatial location of the object, i.e., $o.loc$, is located inside the spatial range of the query $q.MBR$, i.e., 
and when the keywords of the object, i.e., $o.text$, contain all the keywords of the query, i.e., $q.text$.
\end{defn}
\noindent
\textbf{Problem Statement.} In this paper, we study the 
problem of \textit{matching}
an unbounded stream of spatio-textual objects $O$ against a set of continuous spatio-textual queries $Q$.


\begin{figure}
	\centering	\includegraphics[width=2.4in]{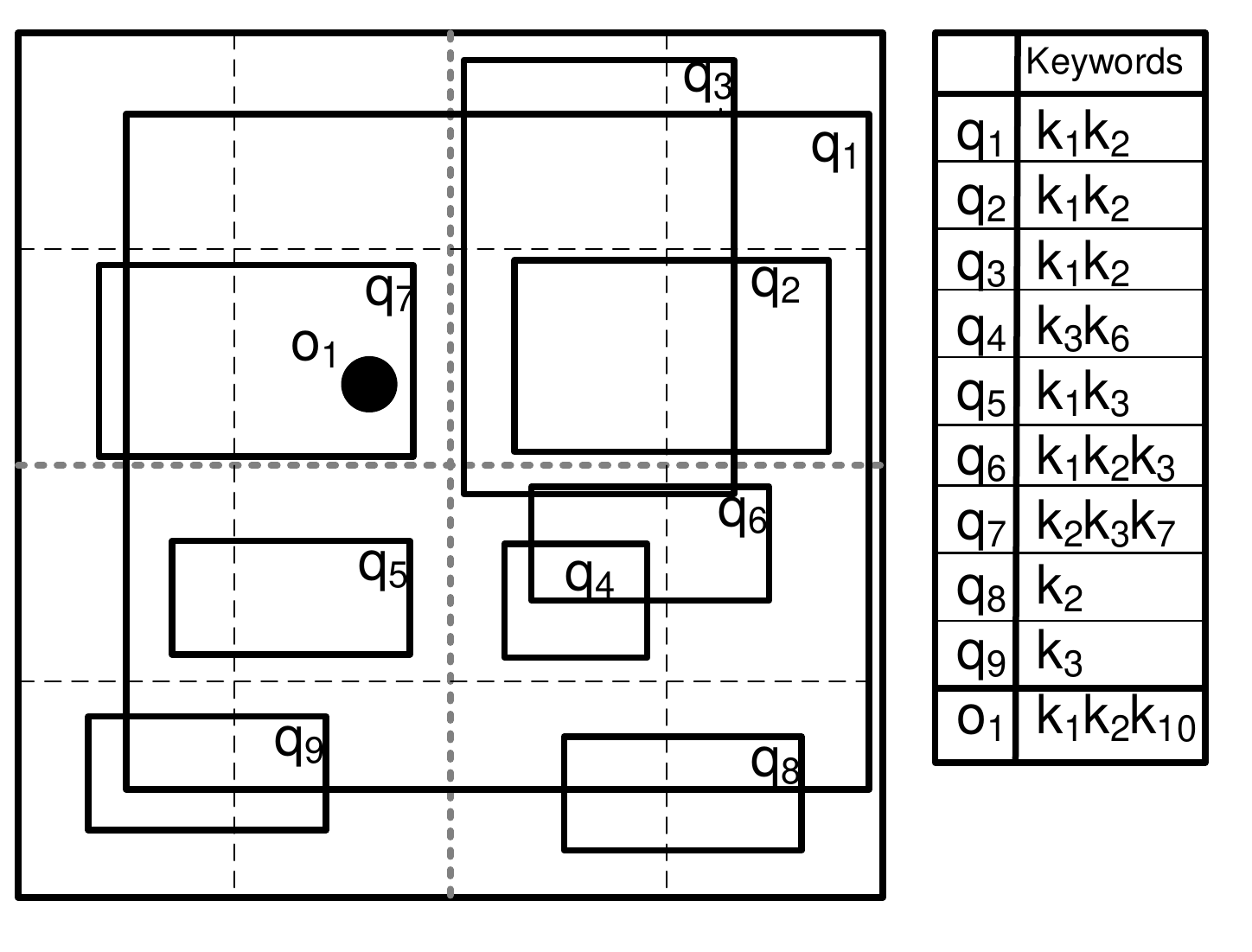}	\caption{Running example.}	\label{fig:runningexample}
\end{figure}

\begin{figure*}
 		\centering
            \subfigure[Ranked-keyword inverted list (RIL)]{	\includegraphics[width=2in]	{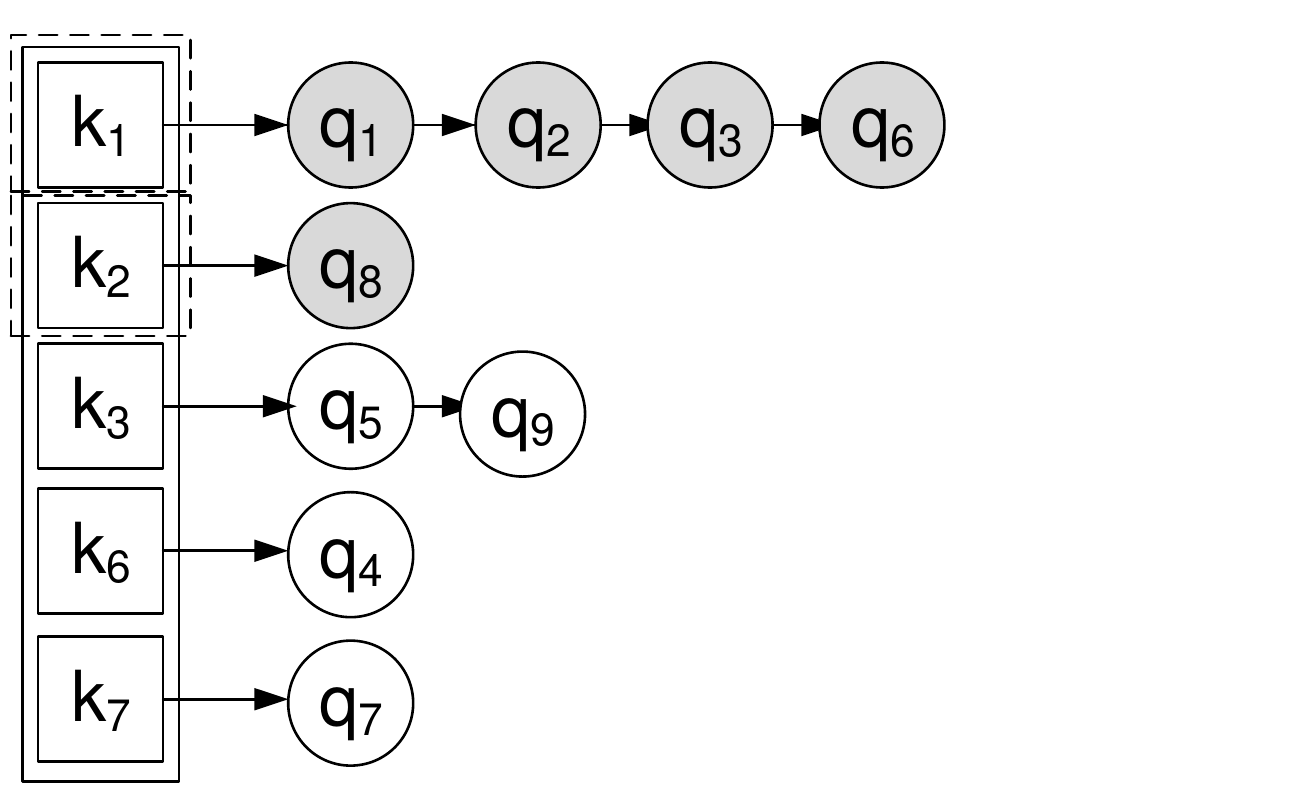}}
         \hfill
        \subfigure[Ordered-keyword trie (OKT)]{	\includegraphics[width=1.7in]	{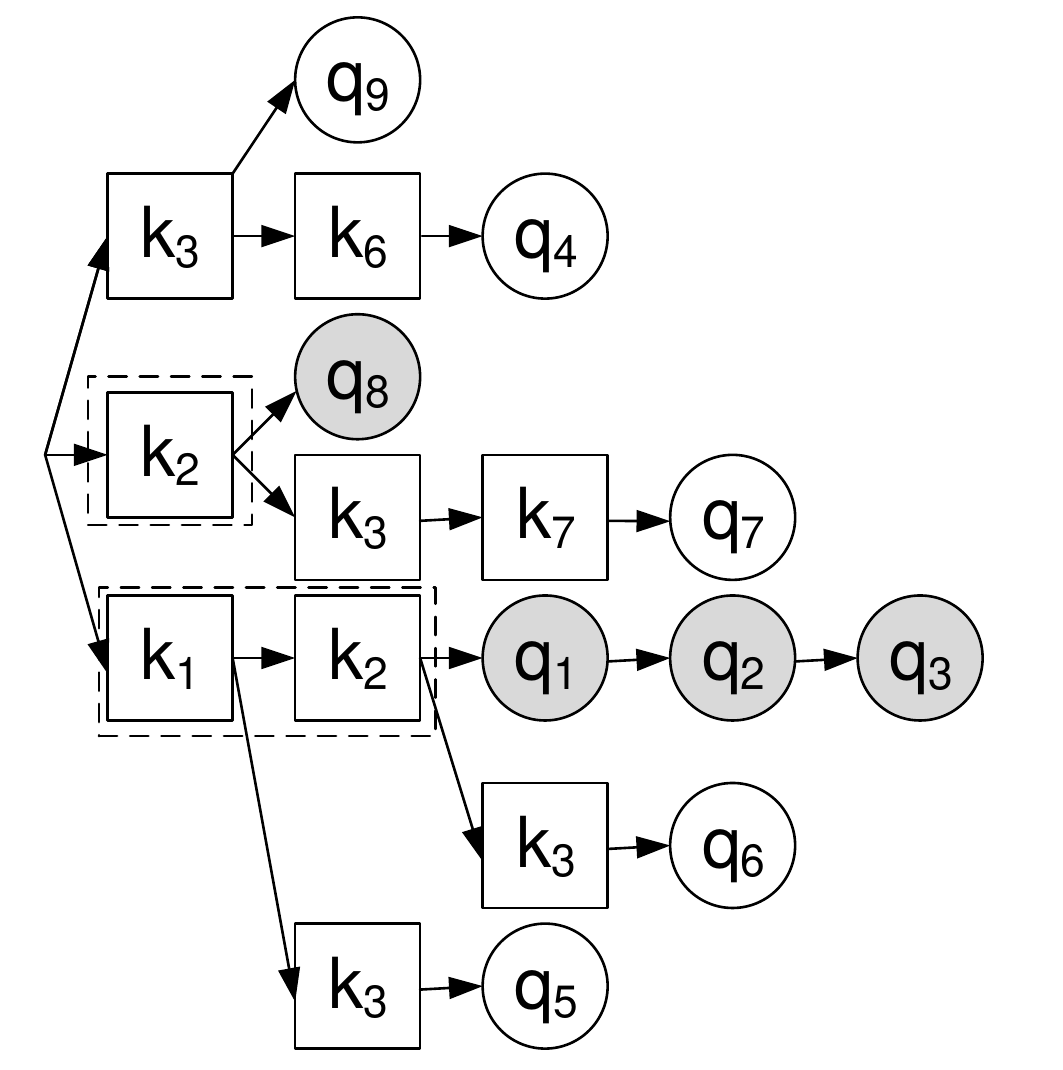}}
        \hfill
        \subfigure[AP-tree]{	\includegraphics[width=2.1in]	{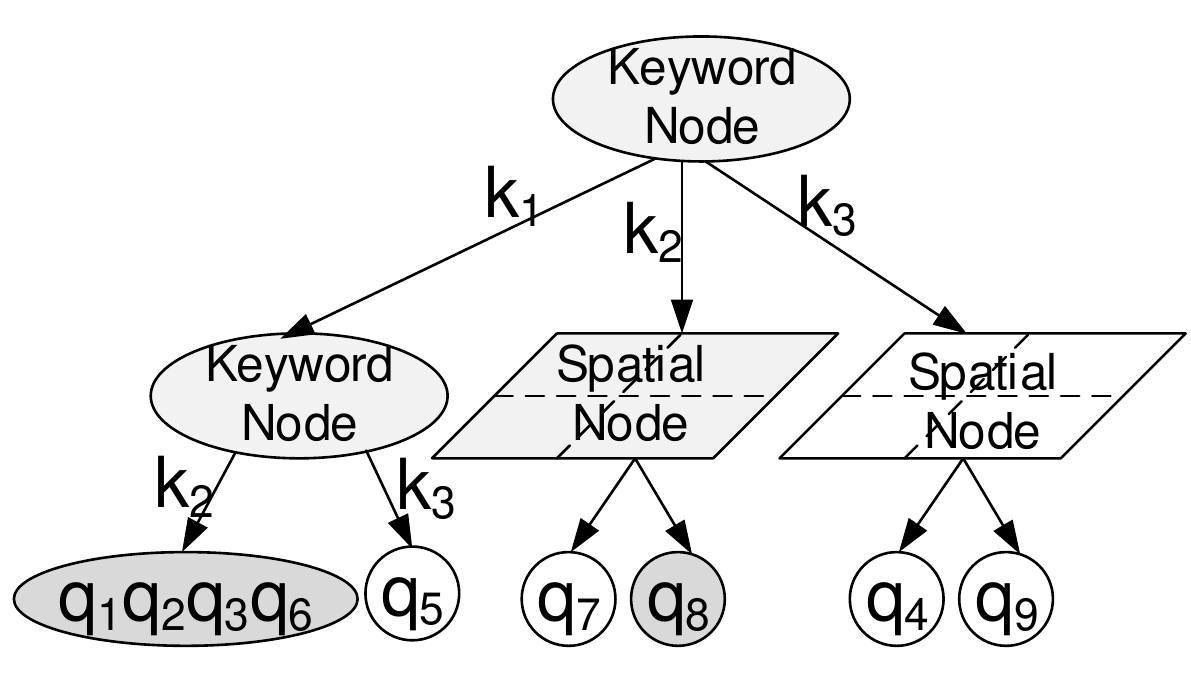}}  
	  \caption{Relevant textual indexes and spatio-textual indexes.   }\label{fig:relevantstructures}
\end{figure*}
\begin{example}\label{example:complexexample}
We use the example given in Figure~\ref{fig:runningexample} throughout the rest of the paper. The figure contains the following nine continuous queries \{$q_1, \cdots, q_9$\}. 
Spatio-textual Object $o_1$ falls inside the spatial range of Queries $q_1$, and $q_7$. However, 
$o_1.text$ fully contains the keywords of only \{$q_1$\}. Thus, $q_1$ is reported as the result of matching $o_1$ against indexed queries.
\end{example}

\subsection{Related Structures}
In FAST, we integrate a spatial index with a 
new
textual indexing approach, 
termed the \textit{adaptive keyword index $AKI$}. To motivate the need for 
AKI, we describe existing textual indexing approaches and discuss their limitations. The two most widely adopted textual indexing approaches are: (1)~the \textit{ranked-keyword inverted list (RIL)}~\cite{zobel2006inverted}, and (2)~the \textit{ordered-keyword trie (OKT)}~\cite{hmedeh2012subscription}. In addition to describing related textual indexes, we  outline the structure of the AP-tree, the state-of-the-art spatio-textual index~\cite{wang2015ap}.\\
\noindent
\textbf{RIL}~\cite{zobel2006inverted} is a data structure for indexing textual items that contain multiple keywords. A spatio-textual filter query, say $q$, can be regarded as a textual item as it contains a set of keywords, i.e., in $q.text$. In RIL, textual items are usually indexed based on their least-frequent keyword. 
Every keyword has a \textit{posting list} of textual items attached to this keyword.
Figure~\ref{fig:relevantstructures}(a) illustrates textual-only indexing of the queries in Example~\ref{example:complexexample} using 
RIL. 
The keywords are usually ranked based on prior knowledge of their frequencies. This imposes a limitation on the efficiency of 
RIL as prior knowledge of the vocabulary of keywords and their frequencies may not be feasible. 
RIL has low memory requirements and has good search performance for objects indexed on infrequent keywords that have short posting lists, e.g., 
as in the posting lists attached to keywords 
$k_6$ and $k_7$. The performance of 
RIL deteriorates when searching for frequent keywords that have long posting lists, e.g., $k_1$. In Figure~\ref{fig:relevantstructures}(a), the posting lists  of the dotted keywords are visited when searching for the keywords of $O_1$ in Example~\ref{example:complexexample}. Because textual items are indexed on a single keyword, search in RIL requires an additional verification step to remove queries whose keywords are not fully contained in the search keywords, e.g., when searching for queries that match the keywords of $o_1$, $q_6$ is initially retrieved as part of the posting list of $k_1$. $q_6$ is removed as $q_6.text\not\subset o_1.text$.\\
\noindent
\textbf{OKT}~\cite{hmedeh2012subscription} is a variation of the traditional trie structure~\cite{knuth1968art} for indexing textual items. The main difference between the traditional trie and 
OKT is that the traditional trie indexes strings using characters 
while
OKT indexes textual objects, e.g., documents, using keywords.  
Figure~\ref{fig:relevantstructures}(b) illustrates textual-only indexing of 
the 
queries in Example~\ref{example:complexexample} using 
OKT. In this figure, keywords are assumed to be ordered lexicographically.  
OKT offers better textual filtering than 
RIL for objects with no infrequent keywords. However, 
OKT has higher memory requirements than 
RIL and does not provide early pruning for indexed objects that contain infrequent keywords, e.g., $k_6$ and $k_7$. Search in OKT follows the traditional trie search algorithm. For example, in Figure~\ref{fig:relevantstructures}(b), the shaded queries attached to dotted keywords are retrieved as the resultset when searching for the keywords of $O_1$ in Example~\ref{example:complexexample}. In contrast to RIL, no additional verification  
is
required when searching OKT as indexing in OKT is based on all the keywords of the indexed item.\\
\noindent
\textbf{The AP-tree~\cite{wang2015ap}} is the current state-of-the-art structure for indexing continuous spatio-textual queries in a streaming environment. When indexing queries, the AP-tree arbitrates between spatial and textual partitioning using an expensive cost function. The AP-tree integrates spatial decomposition 
using
a variation of 
OKT. The main limitations of the AP-tree are: (1)~the AP-tree does not account for the frequencies of keywords to prune queries having infrequent keywords, and (2)~the AP-tree is based on the memory intensive OKT, and has a large memory footprint. Figure~\ref{fig:relevantstructures}(c) illustrates 
how the queries 
in Example~\ref{example:complexexample} 
are indexed 
using the AP-tree. Matching in the AP-tree visits all relevant spatial and textual nodes. Figure~\ref{fig:relevantstructures}(c) illustrates the spatial and the textual nodes, i.e., the shaded nodes, that are visited when matching $O_1$ in Example~\ref{example:complexexample}. The AP-tree requires a verification step to remove non-relevant queries, e.g., $q_6$.

\section{FAST Index Design and Algorithms}\label{sec:structure}


Given the inherent property 
that 
the frequencies of keywords 
follow
a Zipfian distribution, an efficient spatio-textual index needs to account for 
the
frequencies of 
occurrence of the
keywords in real-time and to distinguish between 
the
frequent and infrequent 
ones. 

\begin{figure}[t!]
	\centering \subfigure[Before inserting $q_9$]	{\includegraphics[width=1.6in]{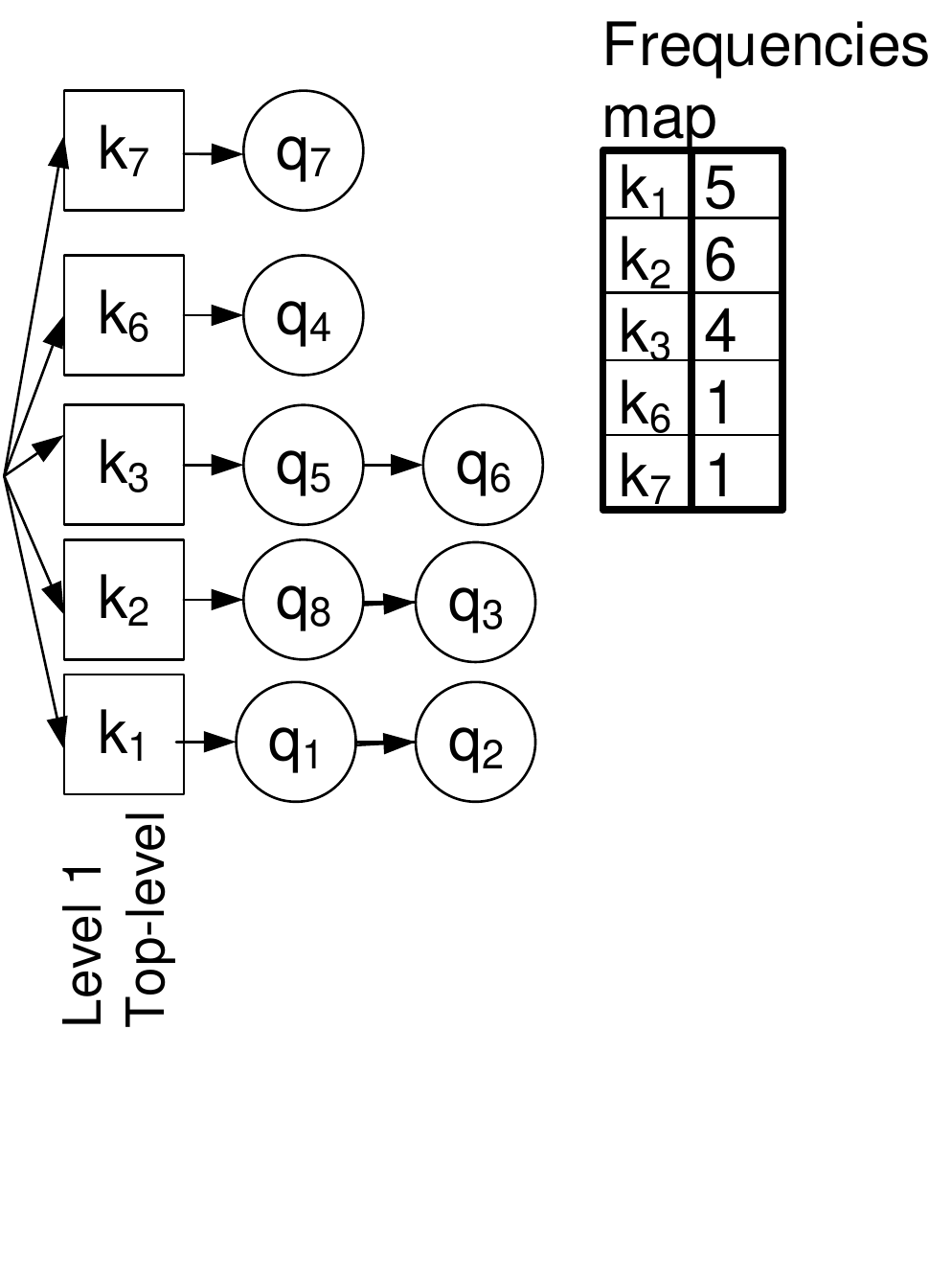}}
    \subfigure[After inserting $q_9$]{	\includegraphics[width=1.6in]{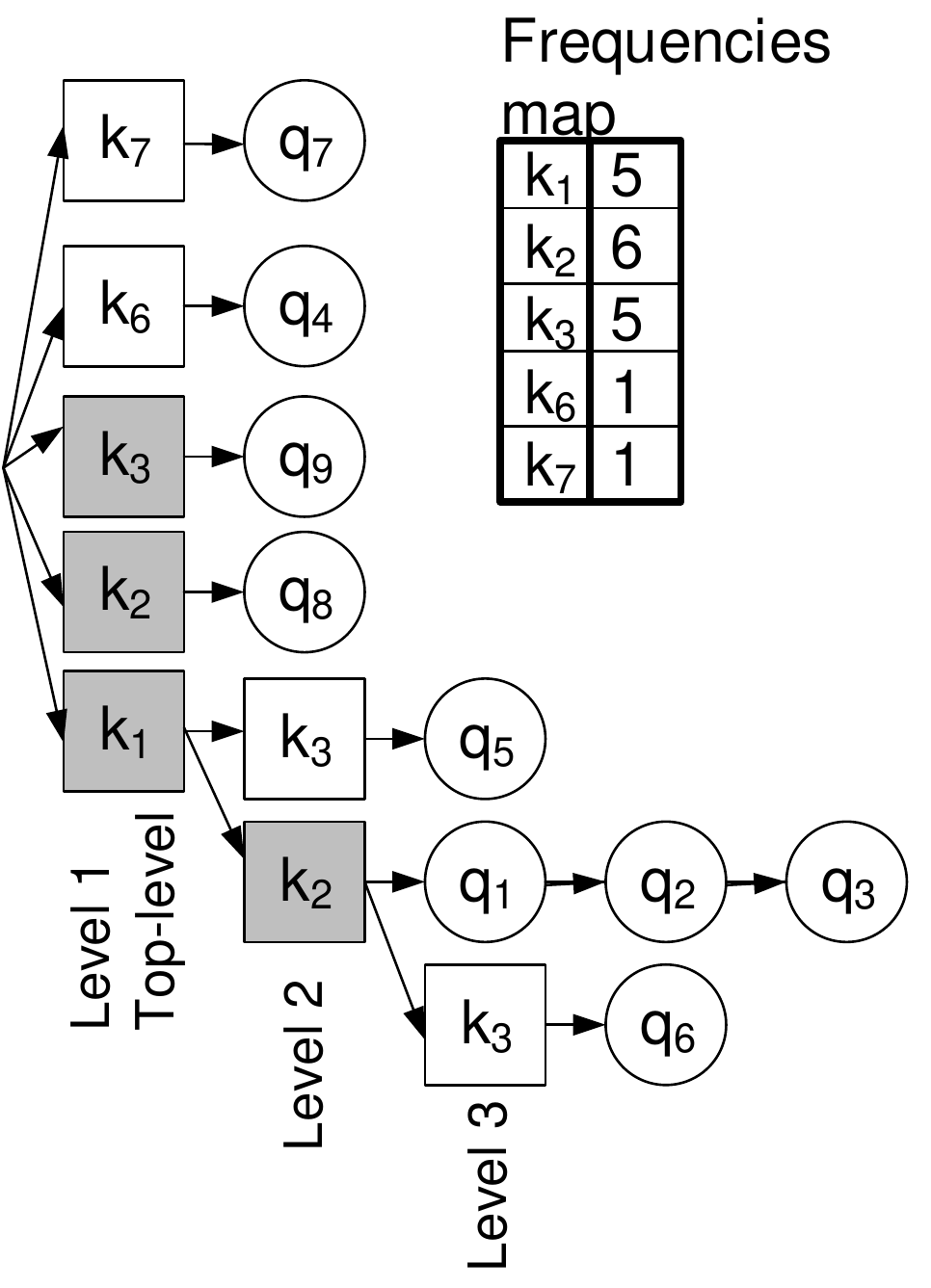}}
    \caption{The adaptive keyword index $AKI$.}	\label{fig:expandablekeyword}
\end{figure}

We equip FAST with a 
new
textual index 
termed, 
the \textit{adaptive keyword index ($AKI$)}. AKI is a text-only index and does not have any spatial discrimination abilities. AKI is integrated with a spatial pyramid to distinguish between queries that are indistinguishable textually. Figure~\ref{fig:expandablekeyword}(b) illustrates an AKI that textually indexes all queries in Example~\ref{example:complexexample}.

AKI is designed as a \textit{multi-level} \textbf{\textit{hash map}} of \textbf{\textit{textual nodes}} with keywords as the key 
to
the hash map (see Figure~\ref{fig:expandablekeyword}).
A textual node, say $N_t$, contains one or both of the following: (1)~a list of queries attached to this node, i.e., $N_t.qlist$, and (2)~a hash map to 
children's
textual nodes with keywords as the key of the hash map, i.e., $N_t.children$. 
Textual nodes are identified using a unique \textbf{\textit{textual-path}} of keywords, e.g., in Figure~\ref{fig:expandablekeyword}(b), Query $q_5$ is attached to 
Textual Node 
[$k_1k_3$], as $q_5$ is stored under the path $k_1,k_3$, where Keywords $k_1$ and $k_3$  are the keywords in $q_5$. 

Textual nodes in AKI are assigned to levels. A \textbf{\textit{top-level}} textual node has no parent textual node and is identified using a textual-path with a single keyword, e.g., in Figure~\ref{fig:expandablekeyword}(b), Textual Nodes [$k_1$], [$k_2$], [$k_3$], [$k_6$], and [$k_7$] are top-level textual nodes. \textbf{ \textit{Leaf}} textual nodes do not have 
child
nodes, e.g., in Figure~\ref{fig:expandablekeyword}(b), Textual nodes [$k_7$], [$k_1k_3$] are leaf textual nodes. Levels of textual nodes in AKI are incrementally numbered, 
where 
the top level is numbered Level 1, as illustrated in Figure~\ref{fig:expandablekeyword}.

Also, for every keyword, say $k_i$, the total number of queries having $k_i\in$  the textual content of queries, is stored in 
a
hash table termed the \textbf{\textit{frequencies map}}. For example, in Figure~\ref{fig:expandablekeyword}(a), the frequencies map indicates that there are five queries containing Keyword $k_1$, i.e., $q_1,\ q_2,\ q_3,\ q_5$ and $q_6$. 

In 
AKI, queries are first indexed to top-level textual nodes using their least-frequent keyword similar to 
RIL.
The least-frequent keyword is identified using the \textbf{ \textit{frequencies map}} and not using prior ranking of 
the
keywords. 
A textual node remains infrequent as long as the number of queries that must be attached to this node in the RIL manner, i.e., the queries do not have any other infrequent keywords to be attached to, is less than a specific threshold, termed the \textit{frequent-keyword threshold} $\theta$. 

\begin{defn}
\textbf{\textit{Frequent-keyword} threshold $\theta$.} The \textit{frequent-keyword} threshold distinguishes between 
the 
infrequent and 
the 
frequent textual nodes. Initially, all textual nodes are infrequent and queries are indexed in the RIL manner, i.e., using a single keyword. When the number of queries that must be attached to an infrequent textual node, say $N_t$, in the RIL manner exceeds $\theta$, $N_t$ is marked as frequent. 
\end{defn}

For example, in Figure~\ref{fig:expandablekeyword}(a), assume that the \textit{frequent-keyword} threshold is two. Before inserting $q_9$, the number of queries attached to all textual nodes is $\leq 2$  and all textual nodes are top-level and are infrequent. $q_9$ has a single keyword $q_3$ and the Textual Node [$k_3$] has two queries attached to it. First, we attempt to transfer some of the queries attached to [$k_3$] to any other infrequent textual node. However, this is infeasible as $q_5$ and $q_6$ only contain keywords $k_1$, $k_2$, and $k_3$ and Textual Nodes 
[$k_1$], [$k_2$], and [$k_3$] have $\theta$ queries attached to them. Hence, Textual Node [$k_3$] is marked as frequent and all queries attached to [$k_3$] get inserted to frequent textual nodes using a lexicographic ordering of their keywords.
We use the lexicographic ordering of keywords as we assume no prior knowledge of the frequencies of keywords, and we cannot use the \textit{frequencies map} to provide a total order on the keywords because values in the \textit{frequencies map} change over time with the insertion and removal of queries.

This requires marking Textual Nodes $k_1$ and $k_2$ as frequent as well. In Figure~\ref{fig:expandablekeyword}(a), queries attached to Textual Nodes [$k_1$], [$k_2$], and [$k_3$] will be re-attached to these textual nodes using the first keyword in their textual content according to a lexicographic ordering of keywords.
In Figure~\ref{fig:expandablekeyword}(b), $q_9$ is attached to [$k_3$]. Also, $q_1,\ q_2,\ q_3,\ q_5,$and $q_6$ should be attached to [$k_1$]. However, the number of queries attached to the \textit{frequent} Textual Node [$k_1$] exceeds $\theta$. The level of [$k_1$] is 1 and AKI uses more keywords to distinguish queries to be attached to [$k_1$] as illustrated in Figure~\ref{fig:expandablekeyword}(b). Textual Nodes [$k_1k_2$] and [$k_1k_3$] are created at Level 2, and Textual Node [$k_1k_2k_3$] is created at Level 3 to distinguish between  $q_1,\ q_2,\ q_3,\ q_5,$ and $q_6$ textually. [$k_1k_3$] is marked as infrequent because only $q_5$ is attached to it.  [$k_1k_2$] is marked as frequent because the textually indistinguishable queries $q_1,\ q_2,$ and $\ q_3$  are attached to it.

Although the number of queries attached to [$k_1k_2$] exceeds $\theta$, these queries are indistinguishable textually and contain exactly the same keywords. Hence, no further discrimination can be performed by the AKI because AKI is a text-only index. AKI lacks any spatial discrimination power, and if we desire to spatially distinguish between queries attached to [$k_1k_2$], we need to integrate spatial pruning abilities with AKI. FAST integrates AKI with a spatial pyramid to combine spatial and textual pruning abilities.

Notice that, in contrast to RIL, AKI attempts to restrict the length of 
the
lists of queries attached to textual nodes to 
prevent 
long lists of queries. However, AKI may contain long lists of queries that are textually \textbf{indistinguishable}. It is a desirable property of textual indexes to group textually indistinguishable
queries. Also, AKI has lower space requirements than 
that of
OKT as AKI requires a lower number of index nodes as 
illustrated
in Figures~\ref{fig:relevantstructures}(b) and~\ref{fig:expandablekeyword}(b).


AKI is adaptive and uses the \textit{frequent-keyword} threshold to create more textual nodes when a higher level of textual discrimination is required. The \textit{frequent-keyword} threshold $\theta$ is very crucial to the performance of FAST. We discuss the experimental tuning of the value of $\theta$, and 
compare the performance 
of
AKI against 
both 
RIL and OKT in Section~\ref{sec:experimentalevaluation}.

\noindent
\textbf{Converting Frequent Textual Nodes to Infrequent Ones.}
AKI keeps track of the frequencies of the keywords of the indexed queries in the \textit{frequencies map}. Whenever a query is removed, the frequencies of the keywords of the removed query are updated in the \textit{frequencies map}. This maintains the dynamic differentiation between frequent and infrequent keywords. Also, updating the \textit{frequencies map} enables converting frequent textual that are no longer frequent, and marking them as infrequent as explained in Section~\ref{subsec:cleaning}.

\noindent
\textbf{Frequency-Aware Spatial Indexing}
Adaptive textual indexing using AKI is insufficient for indexing spatio-textual queries that share the same set of keywords and have different spatial locations. An efficient spatio-textual index needs to 
adapt
to the spatial and textual selectivities of 
the
indexed queries. In Figure~\ref{fig:expandablekeyword}(b), Queries $q_1$, $q_2$, and $q_3$ are attached to the textual node [$k_1k_2$] and cannot be distinguished from each other textually. However, these queries are located at different spatial regions, i.e., can be distinguished from each other spatially. In FAST, we integrate the \textit{spatial pyramid} with 
\textit{AKI} to achieve spatio-textual pruning. The spatial pyramid is a multi-level and a multi-resolution index. Every level in the spatial pyramid contains a spatial grid with a specific granularity. Levels in the spatial pyramid are numbered bottom up and level $0$ is the lowest pyramid level.
\begin{defn}
\textbf{Granularity} at pyramid level $i$ \textbf{\textit{ gran(i):}}\\
is the number of pyramid nodes per dimension at level $i$.
\end{defn}
The top level of the pyramid has a single pyramid node covering the entire indexed space and has a granularity of one. The second level from the top in the spatial pyramid has a granularity of two and contains four cells that covers the entire space.

Let $gran_{max}$ be the maximum supported granularity in FAST. $gran_{max}$ is the pyramid granularity at $level\ 0$, i.e., the lowest pyramid level. The top level in the spatial pyramid is numbered $log_2 (gran_{max})$, e.g., if $gran_{max}$ equals 2, the top level in the spatial pyramid is numbered 1. We discuss the experimental tuning of $gran_{max}$ in Section~\ref{sec:experimentalevaluation}.

We calculate the granularity at level $i$ as follows:
\begin{equation}\label{eq:gran}
gran(i)=\frac{gran_{max}}{2^ii}
\end{equation}
\begin{defn}
\textbf{SideLen(i) }is the side length of a spatial pyramid node at level $i$.
\end{defn}

We define $SideLen_{min}$ as the smallest possible 
side
length size in the spatial pyramid. $SideLen_{min}$ is the side length of spatial pyramid nodes at $level\ 0$, i.e., the lowest spatial pyramid level. We calculate the side length of spatial pyramid nodes at level $i$ as follows:
\begin{equation}\label{eq:step}
SideLen(i)=SideLen_{min} \times (2^{i})
\end{equation}

Every spatial pyramid node within any level, say $i$, has a specific spatial coordinate. To map a spatial location, say $(x_1,y_1)$ into the spatial coordinate $(x_c(i),y_c(i))$ of a pyramid node at Level $i$, we use the following equations:
\begin{equation}\label{eq:cellcorrdinates}
\begin{split}
x_c(i)=\floor*{x_1/SideLen(i)}\\
y_c(i)=\floor*{y_1/SideLen(i)}
\end{split}
\end{equation}

\begin{figure}[t]
	\centering	\includegraphics[width=3.4in]{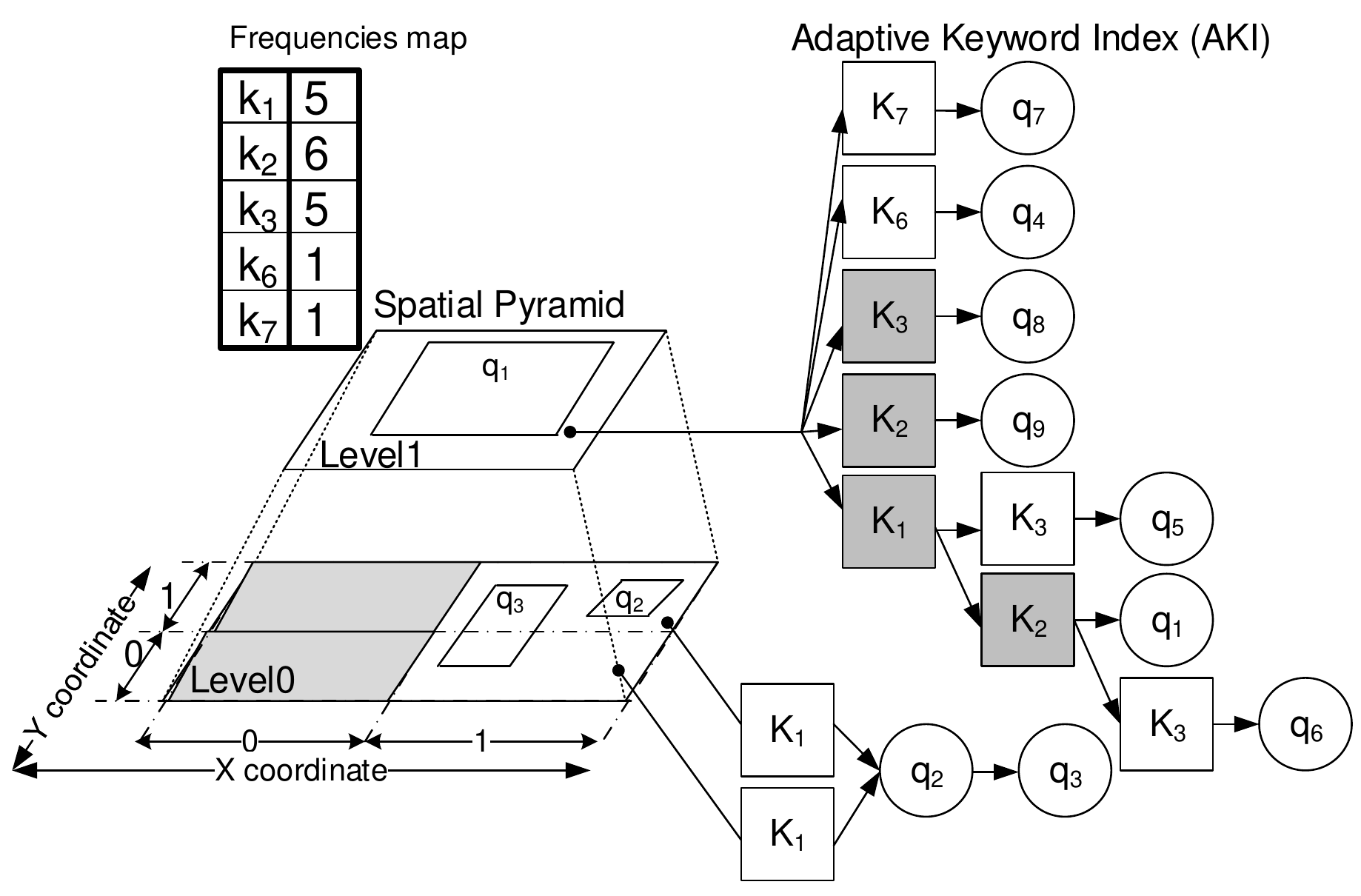}	\caption{The structure of FAST.}	\label{fig:fastindex}
\end{figure}

To reduce the space required by the spatial pyramid, only spatial pyramid nodes that contain queries are instantiated. Empty spatial pyramid nodes are not instantiated and do not consume any memory, e.g., the shaded spatial pyramid node within Level 0 in Figure~\ref{fig:fastindex}. To support this space optimization, all spatial pyramid nodes are accessed using a \textbf{hash table}. The key 
to 
the hash table is the address of the spatial pyramid node. The value is a pointer to the spatial pyramid node.
The address of a spatial pyramid node is calculated using a function of the level number $i$ and the grid \textbf{coordinates} $(x_c,y_c)$ of the spatial pyramid
node as follows:
\begin{equation}\label{eq:nodesaddress}
address(i,x_c,y_c)=i\times gran_{max}^2+y_c\times gran(i)+x_c
\end{equation}
For example, the address of the spatial pyramid node at Level 0 with grid coordinates $(1,0)=0\times 2^2+0\times 2+1=1$.
\begin{figure}[t]
	\centering \subfigure[Without spatial-sharing]	{\includegraphics[width=1.6in]{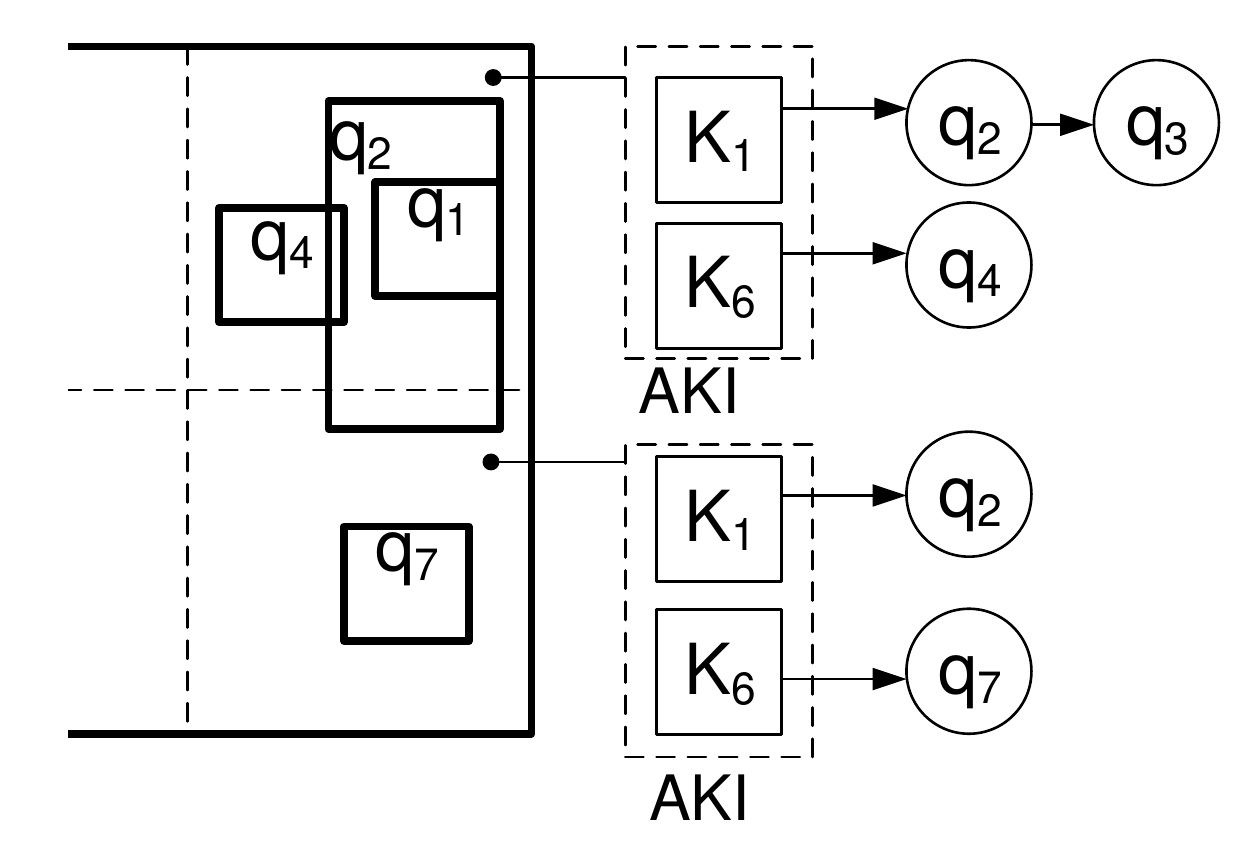}}
    \subfigure[With spatial-sharing]{	\includegraphics[width=1.6in]{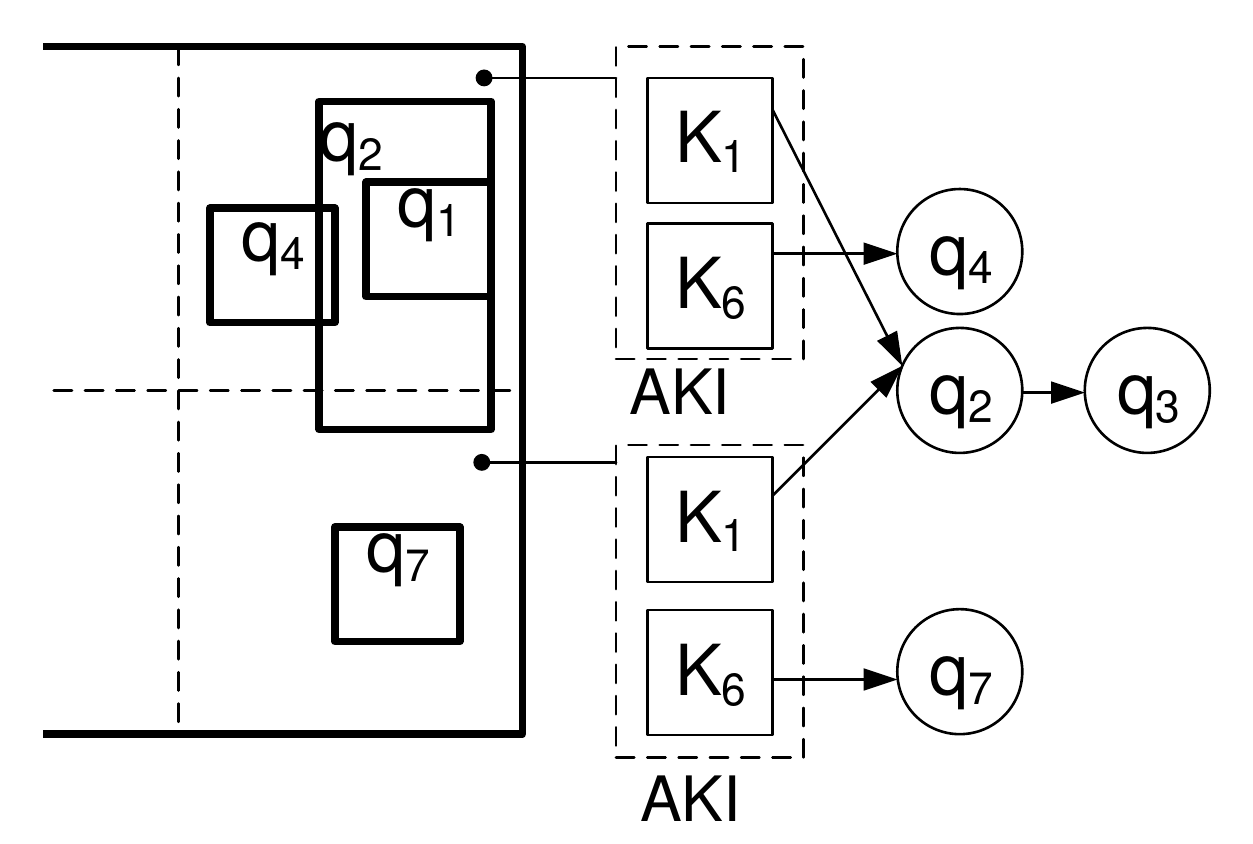}}
    \caption{Spatial-sharing of query lists.}	\label{fig:spatialsharing}
\end{figure}

\noindent
\textbf{Spatial-Sharing of Query Lists}. Each spatial pyramid node contains an AKI instance. To optimize the space required by FAST, we share lists of queries when a query spans multiple spatial pyramid nodes while being attached to infrequent top-level AKI textual nodes. Figure~\ref{fig:spatialsharing}(a) 
illustrates 
two spatial pyramid nodes with two separate AKI indexes. Notice that Query $q_3$ spans 
two 
spatial nodes. In Figure~\ref{fig:spatialsharing}(b), we avoid creating two separate lists of queries to be attached to Keyword $k_1$. We share a single list of queries between two AKI indexes. This reduces the space required for one list of queries. Spatial-sharing of query lists happens at the granularity  of keywords. For example, In Figure~\ref{fig:spatialsharing}(b), query lists attached to Textual Nodes [$k_6$] and [$k_7$] are not shared as these lists do not contain any queries that span more than one spatial pyramid node.

Notice that when a spatial pyramid node, say $N_p$, employs spatial-sharing of query lists  with another spatial pyramid node, say $N'_p$, 
$N_p$
may 
also 
point to a few extra queries that overlap only with Node $N'_p$. For example, in Figure~\ref{fig:spatialsharing}(b), Query $q_1$ is attached to both spatial pyramid nodes. However, $q_1$ spans only the top spatial pyramid node. This is acceptable and does not introduce overhead in the matching processing as the length of 
the 
shared lists is restricted to
the 
\textit{frequent-keyword} threshold $\theta$. Before marking an AKI textual node as frequent, we check if a spatially-shared query list is attached to the AKI textual node. We separate the spatially-shared lists and remove non-spatially overlapping queries to reduce the length of $qlist$
that is
attached to the AKI textual node to prevent unnecessary marking of textual nodes as frequent. \\
\noindent
\textbf{Frequency-Aware Spatio-textual Indexing}.
Initially, all queries are indexed at the top level of the spatial pyramid. When the number of queries attached to a frequent AKI textual node exceeds a specific threshold, e.g., Textual Node [$k_1$$k_2$] in Figure~\ref{fig:expandablekeyword}(b), queries are partitioned to descend to a lower spatial pyramid level, i.e., to a spatial pyramid level with higher resolution. A spatial pyramid node in any spatial pyramid level, say $i$, other than Level 0, potentially covers four children spatial nodes in the spatial pyramid level directly below $i$, i.e., $i-1$. When the number of queries attached to a frequent textual node exceeds $4 \theta$, the queries are sorted based on their ranges. Queries having area less than the median of the sorted 
query
list descend to a lower pyramid level. Queries with smaller ranges are chosen to descend as these queries have higher probability of joining different spatial nodes at the lower pyramid level. This adaptively captures the difference in frequencies across different spatial regions.

Figure~\ref{fig:fastindex} illustrates the hybrid structure of FAST. In this figure, a two-level spatial pyramid is integrated with AKI. Assume that the \textit{frequent-keyword} threshold is two. For the sake of illustration, assume that queries descend when the number of queries attached to a frequent textual node exceed $1\times \theta$ instead of $4\times \theta$. In Figure~\ref{fig:expandablekeyword}(b),  
$q_1$, $q_2$,
and $q_3$ are all attached to 
Textual Node
[$k_1k_2$]. This calls for a 
descent 
of queries to improve spatial discrimination. $q_1$, $q_2$, and $q_3$ are sorted according to the area of their spatial ranges. $q_1$ remains in 
Level 1 of
the spatial pyramid while 
$q_2$ and $q_3$ 
descend to Level 0. Also, Queries
$q_2$ and $q_3$
are shared between two infrequent AKI textual nodes at Level 0 of the spatial pyramid. 

Notice that only queries with no infrequent keywords at 
level $i$
descend to level $i-1$. For example, $q_4$ has Keyword $k_6\in q_4.text$. No other query contains $k_6$, and $k_6$ is an infrequent 
keyword
and the number of queries attached to 
$k_6$
is less than $\theta$. Hence, all queries containing $k_6$ remain at Level 1, and 
there 
is no need to search for $k_6$ in Level 0. This improves the matching performance by reducing the number of keywords being searched for as the search goes down the spatial pyramid.

Notice that when queries descend the spatial pyramid, they may be replicated to more than one pyramid node with increased memory overhead. The replication overhead becomes more significant when queries with large spatial ranges descend to lower pyramid-levels with higher resolution because these queries will span 
multiple
pyramid nodes. 
As a heuristic, 
to reduce the number of queries
with large spatial ranges 
that descend to lower pyramid levels with higher resolution, we set the lowest spatial pyramid level a query can descend 
into to be 
the level having a slide length that is strictly greater than the side length of the query. 
We refer to the side length of Query $q$ by $q.r$, where $q.r$ is calculated as follows:
\begin{equation}~\label{eq:sidelength}
q.r=max((q.x_{max}-q.x_{min}),(q.y_{max}-q.y_{min}))
\end{equation}
The lowest level $L_{min}$ of Query $q$ is calculated as follows:
\begin{equation}
L_{min}(q)=\ceil{\log_{2}(\floor{\dfrac{q.r}{SideLen_{min}}})}
\end{equation}

We analyze 
the replication of queries in Appendix~\ref{app:replication}.
\subsection{Algorithms}\label{subsec:algorithms}
In this section, we present the indexing, 
searching, and cleaning 
algorithms of FAST.
\begin{algorithm}[h]
\setstretch{.85}
update $frequenciesMap$ for $q.text$\\
$key_{minfreq} \gets  getLeastFrequentKeyword(q.text)$\\
$PN_{list} \gets getRelevantPyarmidNodes$\\
$sharedList \gets null$\\
\ForEach{Pyramid Node $N_P$ in $PN_{list}$}{
$queue \gets \{\}$ \\
	$N_t \gets N_P.get(key_{minfreq}$)\\
	\If{$sharedList$ is not null and ($N_t.qlist$ is infrequent and |$N_t.qlist$|+|$sharedList$|<$\theta$))}{
    	$N_t.qlist\gets sharedList\gets Merge(N_t.qlist,sharedList)$ 
    }
    \ElseIf{$N_t$ is infrequent }{
    	$N_t.qlist.add(q)$\\
        \If{$|N_t.qlist|\leq \theta$}{
        	$sharedList\gets N_t.qlist$ \\
        }\Else{
        	$queue.addAll(N_t.qlist)$\\
            mark $N_t$ as frequent\\
        }
    }
    \ForAll{Query $q_e$ in $queue$}{
    	\If{$q_e$ can be inserted to another infrequent textual node}{
    		insert $q_e$ into the other textual node
   		 }\Else ( \tcp{insert $q_e$ to the appropriate frequent textual node based on lexicographic order of the keywords in $q_e.text$ }){
     	i=1\\
    	$N_t\gets N_P.get(q_e.text[i])$\\
    	\While{i$N_t$ is frequent and $i\leq |q_e.text|$}{
        	$i\gets i+1$;\\
            $N_t\gets N_t.children.get(q_e.text[i])$
         }
         $N_t.qlist.add(q_e)$\\
         \If{$|N_t.qlist|>\theta$ and $N_t$ is infrequent}{
            Mark $N_t$ as frequent\\
            Split $N_t.qlist$ into the subsequent textual level using one more keyword 
         } \ElseIf {$|N_t.qlist|>4\theta$ and $N_t$ is frequent }{
            dlist.add(queries to descent in  $N_t.qlist$)\\
        }
    }    
 }
 \ForAll{Query $q_d$ in $dlist$}{
            Insert($q_d$,level-1)
  }
}
\caption{Insert( $q$,level)}\label{alg:queryinsert}
\end{algorithm}

\subsubsection{Insertion Algorithm}
First, we update the frequencies of keywords in the 
\textit{\textbf{frequencies map}} according to the query being inserted. 
Then, at the top level of the spatial pyramid, we attempt to attach the incoming query to an infrequent AKI textual node using 
 the
 least-frequent 
 keyword of the query, i.e., $key_{minfreq}$. 
If the query has more than one infrequent keyword, $key_{minfreq}$ is chosen arbitrarily from the set of keywords with the minimum frequency. 
If the incoming query cannot be attached to any infrequent AKI textual node, we index the query to frequent AKI nodes according to the lexicographic ordering of the keywords of the query. After attaching the incoming query to a frequent AKI textual node, we check if a descend operation is required, i.e., when the number of queries attached to the frequent AKI textual node exceeds $4\times \theta$. If a descend operation is required, we identify queries to be descended, and we recursively reinsert queries these queries at the subsequent level of the spatial pyramid.

If the incoming query spans more than one spatial pyramid node and the incoming query is attached to infrequent AKI textual nodes in all relevant spatial pyramid nodes, then we employ 
 \textit{spatial-sharing of query lists} to share the textual index among the spatial nodes to reduce the memory footprint of FAST.  Algorithm~\ref{alg:queryinsert} describes the query indexing algorithm adopted in  FAST.


\begin{algorithm}[t]
\setstretch{.85}
$keywords \gets o.text$\\
\For{$level=level_{max};level>=level_{min};level--$}{
	nextLevelKeywords $\gets$ \{\}\\
	$N_p  \gets getSpatialPyramidNode(level,o.loc)$\\
    \If{$N_p$ is not Null}{
      \For{$i=1;i<=|keywords|;i++$}{
      	$N_t \gets N_p.get(keywords[i])$\\
        \If{$N_t$ is infrequent}{
        	\ForEach {Query q in $N_t.qlist$ }{
              \If{$q$ not expired and $o.loc$ inside $q.MBR$ and $keywords$ contains $q.text$   }{
              	add q to result\\
              }
    		}
        }\Else{
        	nextLevelKeywords.add(keywords[i])\\
        	SearchFrequent($N_t,i,o,keywords$)
        }
      }
    }
    keywords $\gets$ nextLevelKeywords\\
}
\caption{Match(Data Object $o$)}\label{alg:objectmatching}
\end{algorithm}
\begin{algorithm}[t]
\setstretch{.85}
\If{$N_t$ is infrequent}{
\ForEach {Query q in $N_t.qlist$ }{
              \If{$q$ not expired and $o.loc$ inside $q.MBR$ and $keywords$ contains $q.text$ }{
              	add q to result\\
              }
    		}
 }\Else{
	\ForEach {Query q in $N_t.qList$}{
      \If{$q$ not expired and $o.loc$ inside $q.MBR$ }{
      add q to result\\
      }
    }
    \For{$j=(i+1);j\leq |keywords|;j++ $}{
    	SearchFrequent($kIndex.children.get(keywords[j]),$\\$j,keywords$)
        }
    }

\caption{SearchFrequent($N_t,i,o,keywords$)}\label{alg:searchexapandable}
\end{algorithm}
\begin{algorithm}[h]
\setstretch{.85}
$N_p\gets$ cleaningQueue.dequeue()\\
 \ForEach {Textual Node $N_t$ in $N_p$}{
    \ForEach {Query q in $N_t.qlist$}{
    	\If{$q.t_{exp}< currenttime$}
        {
        	$N_t.qlist$.remove($q$)\\
            \If{$q$ not marked as deleted}{
            	mark $q$ as deleted\\
                 \ForEach {Keyword k in $q.text$}{
                 	frequenciesMap(k)--\\
                    \If{frequenciesMap(k)==0}{
                    	$N_p.remove(k)$
                    }    
                 }
            }     
        }
    }
}
\If{$N_p$ has no textual nodes}{
    FAST.remove($N_p$)
}\Else{
	cleaningQueue.enqueue($N_p$)
}
\caption{Clean} \label{alg:clean}
\end{algorithm}
\subsubsection{Matching Algorithm}
When an incoming data object arrives, 
it needs to be inspected
against relevant 
pyramid and AKI nodes. 
Matching in FAST consists of the following three steps: (1)~Identify
relevant pyramid nodes, (2)~Search AKIs within 
the
relevant pyramid nodes, and (3)~Refine 
the
results to remove 
the
expired queries. Algorithm~\ref{alg:objectmatching} describes the matching algorithm adopted in  FAST. 
The matching process starts from the highest pyramid-level. For a data object, say $O$, with point spatial location $O.loc$, at most one pyramid node per level is relevant for matching. 
The data object that has a point location cannot overlap more than one spatial pyramid node per level because there is no overlap in spatial ranges of pyramid nodes in the same level.
We calculate
the index of every relevant pyramid node using ~\Crefrange{eq:step}{eq:nodesaddress}. 
We assume that keywords of the data objects are sorted lexicographically.  
We retrieve textual nodes for every keyword in $O.text$, where $O.text$ is the set of keywords of Data object $O$ that is being matched against FAST. 
If the top-level 
textual node, say $N_t$, 
is infrequent, we verify the spatial and textual criteria of all queries in $N_t.qlist$. If Node $N_t$ is frequent, children of this node are recursively searched as outlined in Algorithm~\ref{alg:searchexapandable}. 
In the matching process, spatial validation of queries verifies that the data object is located inside the spatial range of the query. Textual validation verifies 
that this
data object
contains all the keywords of the query.

Notice that
queries directly attached to frequent textual nodes do not require textual validation  as these queries only contain the keywords that constitute the path of the frequent textual node.
For example, in 
Figure~\ref{fig:fastindex}, consider 
Query $q_1$ 
that is attached to the frequent textual node [$k_1k_2$] in Pyramid Level 1. This query has only two keywords 
$k_1$ and $k_2$.
If more keywords 
are to exist
in $q_1$, $q_1$ would have been attached to a child node of [$k_1k_2$]. However, queries attached to infrequent textual nodes 
require
textual validation as these queries may contain more keywords than the path of the textual node. For example, in Figure~\ref{fig:fastindex}, Query $q_4$ has more keywords, i.e., $k_3$, than the path of the infrequent textual node [$k_6$] in Level 1, and hence requires additional textual validation at matching time.

Notice that in FAST, Keywords being searched 
in Pyramid Level i-1 are a subset of keywords being searched 
in Level i $\subseteq$ $o.text$, where $o$ is the spatio-textual object being matched. Recall that matching in FAST is top-down and the lowest pyramid level in FAST is Level $0$. All queries attached to an infrequent top-level textual node in 
Level $i$
can never descend to be indexed at Level $i-1$ for the same spatial range.
All infrequent top-level textual nodes 
at a pyramid node, say $N_p$, 
within Level $i$, correspond to a set of keywords, say $SU_i$. 
$SU_i$ can never exist at a 
pyramid node, say $N'_p$,
at Level $i-1$ that shares the same spatial range with $N_p$. Hence, 
at matching time,
the Set $SU_i$ is not considered for matching at Level $i-1$.

The final step in the matching process is to remove 
the expired queries 
from the resultset and to verify the spatial overlap between the incoming data object and 
the
matched queries.

\subsubsection{Index Maintenance}\label{subsec:cleaning}
Over time,
some indexed queries expire, 
and some 
new queries 
get inserted.
FAST employs a lazy 
vacuum-cleaning mechanism that maintains the structure of FAST and updates the frequencies of 
the 
keywords of 
the
indexed queries. 
The vacuum cleaner has the following functionalities: (1)~Detect and remove 
the
expired continuous queries, and (2)~Reflect the current frequencies of keywords according to the 
expired and removed queries.
Algorithm~\ref{alg:clean} describes the cleaning procedure used in FAST.  

The vacuum cleaner maintains a  queue of the pyramid nodes. 
In every
\textit{cleaning interval} $I$, the vacuum cleaner visits a pyramid node to be cleaned, 
say $N_p$, from
the top of the cleaning queue. 
Then, the cleaner 
iterates over all textual nodes within $N_p$ and scans all 
the
queries attached to 
the
textual nodes within $N_p$ to check for 
the expired queries,
i.e., $q.t_{exp} < current_time$. The vacuum cleaner updates the frequencies of keywords of a removed query. The vacuum cleaner needs to account for 
the
expired queries that span multiple pyramid nodes to 
avoid updating
the frequencies multiple times. When an expired query is first removed, the 
vacuum cleaner
updates
the keyword statistics and marks the expired query to prevent updating the statistics more than once.   

When the frequency of 
a
keyword in the \textit{frequencies map} 
reaches
zero, 
the
textual nodes associated with this 
keyword
are removed.  When 
an entire
pyramid node 
becomes empty,
the entire node is removed from FAST. Notice that in the lazy cleaning approach, expired continuous queries are not removed instantaneously. Instead, they may remain indexed until the vacuum cleaner touches them. However, this does not affect the correctness of matching in FAST because the matching algorithm in FAST has a refinement step that removes expired queries from the matching result.\\
\noindent
\textbf{Indexing Queries with General Boolean Expressions.}
FAST supports matching data objects against queries whose keywords are fully contained in the keywords of the incoming data object. Also, FAST is able to support queries with general boolean expressions on their keywords. For example, the textual condition of a query, say $q$, is to be matched against all data objects that either contains ($k_1$ and $k_2$) or ($k_3$ and $k_4$). This textual condition is a boolean expression in the disjunctive normal form (DNF). We address boolean expressions in DNF because boolean expressions represented in the conjunctive normal form (CNF) can be converted to DNF~\cite{enderton2001mathematical}. To support queries in DNF, we instantiate a sub-query per conjunction. For example, $q$ is split into two sub-queries $q_1$ and $q_2$, where $q_1.text$ is $\{k1,k2\}$ and $q2.text$ is $\{k_3,k_4\}$. Sub-queries $q_1$ and $q_2$ have pointers to $q$. Then, $q_1$ and $q_2$ are indexed using the insertion algorithm of FAST. If a sub-query, e.g., $q_1$, appears in the matching resultset, the original query, i.e., $q$, is reported in the final resultset. To avoid duplicate results when more than one sub-query qualifies in the matching process, a flag is added to the original query when it is first added to the matching resultset. This flag is cleared at the end of the matching process.\\
\noindent
\textbf{Matching Objects with Rectangular Spatial Ranges.}
FAST supports matching data objects with point location. Also, FAST is able to support the matching of data objects with rectangular spatial locations. Matching of rectangle data objects starts from the top level of the spatial pyramid. When the matching algorithm 
descends
the spatial pyramid, the matching algorithm visits all the nodes of the spatial pyramid that overlap the rectangular range of the incoming data object. When a query, say $q$, spans multiple spatial nodes that overlap the rectangular spatial location of an object being matched, $q$ may appear multiple times in the matching resultset. The matching algorithm prevents duplicate results by adding a flag to mark queries added to the matching resultset. This flag is cleared at the end of the matching process.

\subsection{Analysis}\label{subsec:analysis}
In this section, we analyze the matching time for FAST. 
AKI within FAST is proposed to address the limitations of existing textual indexes, i.e., the deterioration in the
matching performance in RIL, and the large memory requirements of OKT. 
RIL's matching performance deteriorates due to the existence of long posting lists of indexed objects that have no infrequent keywords. OKT has an advantage over RIL in the matching performance as OKT uses multi-level indexing that uses all the keywords of an indexed object. However, this increases the memory footprint of OKT. AKI  uses the \textit{frequent-keyword} threshold $\theta$ to restrict the number of queries attached to textual nodes. This creates balance between the memory requirements and the matching performance in AKI.  We measure the matching performance of an index by the number of index nodes visited during matching.

To analyze the matching performance of AKI, we first study the matching performance ($MP$, for short) of 
RIL for a set of keywords $S$.  The total number of textual items visited when matching $S$ against the indexed objects is
\begin{equation}\label{eq:ril}
MP_{RIL}(S)= \sum_{j=1}^{|S|}|RIL[s_i]|
\end{equation}
where $|S|$ is the number of keywords being searched in 
$S$, and $|RIL[s_i]|$ is the number of indexed textual items attached to the Keyword $s_i$. 

OKT is a multi-level keyword index that is illustrated in Figure~\ref{fig:relevantstructures}(b). The matching process of the keyword Set $S$ at level $i$ 
in
OKT iterates over the keywords in $S$ to find a subset of matching keywords to proceed to level $i+1$ in 
$OKT$ 
(Notice that, in OKT, level numbers increase as we descend the index).
OKT assumes a total order of the indexed keywords. 
For a matched keyword,
say  $s_j$,  at level $i$ of 
$OKT$, 
the search proceeds to level $i+1$ with the keyword set [$S$-$\{\ s_1,\ s_2,\ \dots,\ s_j\}$]. 
Hence, at level $i$, the matching time for $OKT$
can be expressed recursively as follows~\cite{hmedeh2012subscription}:\\
$MP_{OKT}(i,S)$=
\begin{equation}\label{eq:okt}
|S|+\sum_{j=1}^{|S|} \alpha_{ij} \times MP_{OKT}(i+1,S-\{s_1,\dots,\ s_j \})
\end{equation}
where 
$\alpha_{ij}$ is the probability of having Keyword $S_j$ indexed at level $i$.  $\alpha_{ij}$ 
depends on the 
frequencies of 
the
indexed keywords and their 
probabilities
of co-occurrence. The recursion in Equation~\ref{eq:okt} terminates at the deepest level of OKT, i.e., the largest $|q.text|$ for any indexed query $q$. Notice that Equation~\ref{eq:okt} is a recurrence relation and is not in closed form. However, for datasets with known probabilities
of keyword co-occurrence and a bounded $|q.text|$ for any indexed query $q$, a closed formula can be devised. For textual items with infrequent keywords, AKI has a similar behavior to 
that of
$RIL$, yet with a restricted length of posting lists, i.e., $\theta$.  For textual items with no infrequent keywords, AKI has a similar behavior to 
that of
$OKT$. From Equations~\ref{eq:ril} and~\ref{eq:okt}, we estimate the matching performance of AKI as follows:\\
$MP_{AKI}(i,S)$= 
\begin{equation}\label{eq:eki}
 \begin{cases}
   |S|\times  \theta, infrequent\\
   |S|+\sum_{j=1}^{|S|} \alpha_{ij} MP_{AKI}(i+1,S-\{s_1,\dots,s_j \}), frequent
  \end{cases}
\end{equation}
Similar to Equation~\ref{eq:okt}, the recursion in Equation~\ref{eq:eki} terminates at the deepest level of AKI.\\
\noindent
\textbf{Estimation of The \textit{Frequent-Keyword} Threshold.}
We use Equation~\ref{eq:eki} to estimate an upper bound on $\theta$. The  matching performance of infrequent AKI nodes should not exceed the matching performance of frequent AKI nodes. In the worst case, frequent AKI nodes resemble an OKT index. 
\begin{equation}\label{eq:theta}
\theta \leq \dfrac{MP_{OKT}}{|S|}
\end{equation}
From Equation~\ref{eq:eki}, the 
worst-case
matching in FAST requires AKI matching at every level of the spatial pyramid. The matching performance in FAST can be estimated as follows:
\begin{equation}\label{eq:fastmatching}
MP_{FAST}(S)= log(gran_{max})\times MP_{AKI}(0,S)
\end{equation}
where 
$log(gran_{max})$ represents the height of the spatial pyramid.



\section{Experimental Evaluation}\label{sec:experimentalevaluation}
In this section, we compare the performance of FAST against the performance of the state-of-the-art index, 
the AP-tree~\cite{wang2015ap}. 

\subsection{Experimental Setup}
\begin{table}[!t]
\centering
    \caption{
    The datasets used in the experiments.}     
    \label{tab:dataset}
    \begin{small}
    \begin{tabular}{|l|l|l|l|}
    \hline
    {\bfseries Dataset} & {\bfseries Tweets} & {\bfseries Synthetic} & {\bfseries Places} \\
    \hline
   Number of entries    &30M&30M&12.9M	\\
    \hline
    Vocabulary size      &804K&804K&854k	 \\
        \hline
      Avg num of keywords/Entry     &4&4&9	 \\
        \hline
    \end{tabular}
    \end{small} 
\end{table}
\begin{table}[!t]
\centering
    \caption{The values of the parameters used in the experimental evaluation.}     
    \label{tab:experiments}
    \begin{small}
    \begin{tabular}{|l|l|l|l|}
    \hline
    {\bfseries Parameter} & {\bfseries Value} \\
    \hline
    Number of queries (million)       &1,2.5,~\textbf{5},7.5,10,20 	\\
    \hline
    Number of query keywords       &1, 2,~\textbf{3}, 5, 7	 \\
    \hline
    Spatial side-length of a query &.01\%,.05\%,.1\%,.5\%,\textbf{1\%},5\%,10\%	\\
        \hline
    \end{tabular}
    \end{small} 
\end{table}
\begin{figure*}
 	\centering
\begin{minipage}{.8\textwidth}
\raggedleft
            \subfigure[Matching time AKI]{	\includegraphics[width=1.3in]	{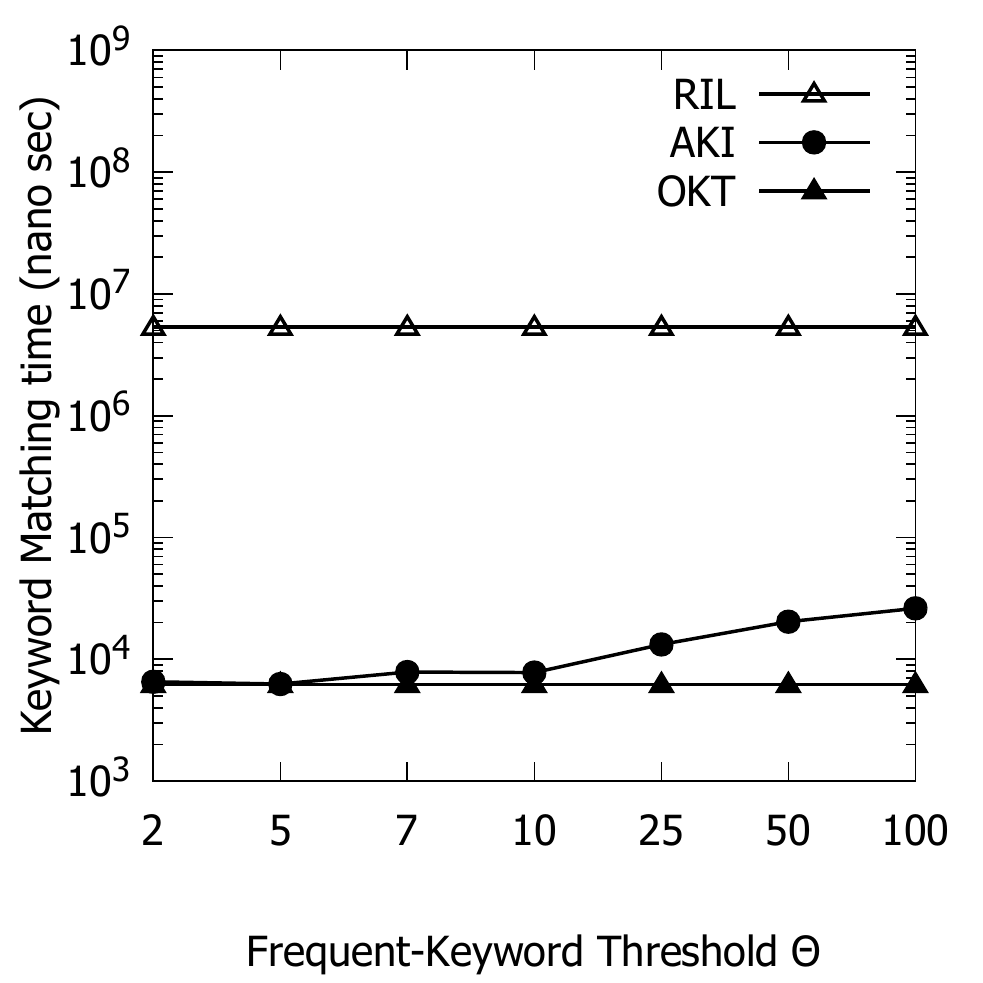}}
        \subfigure[Memory footprint AKI]{	\includegraphics[width=1.3in]	{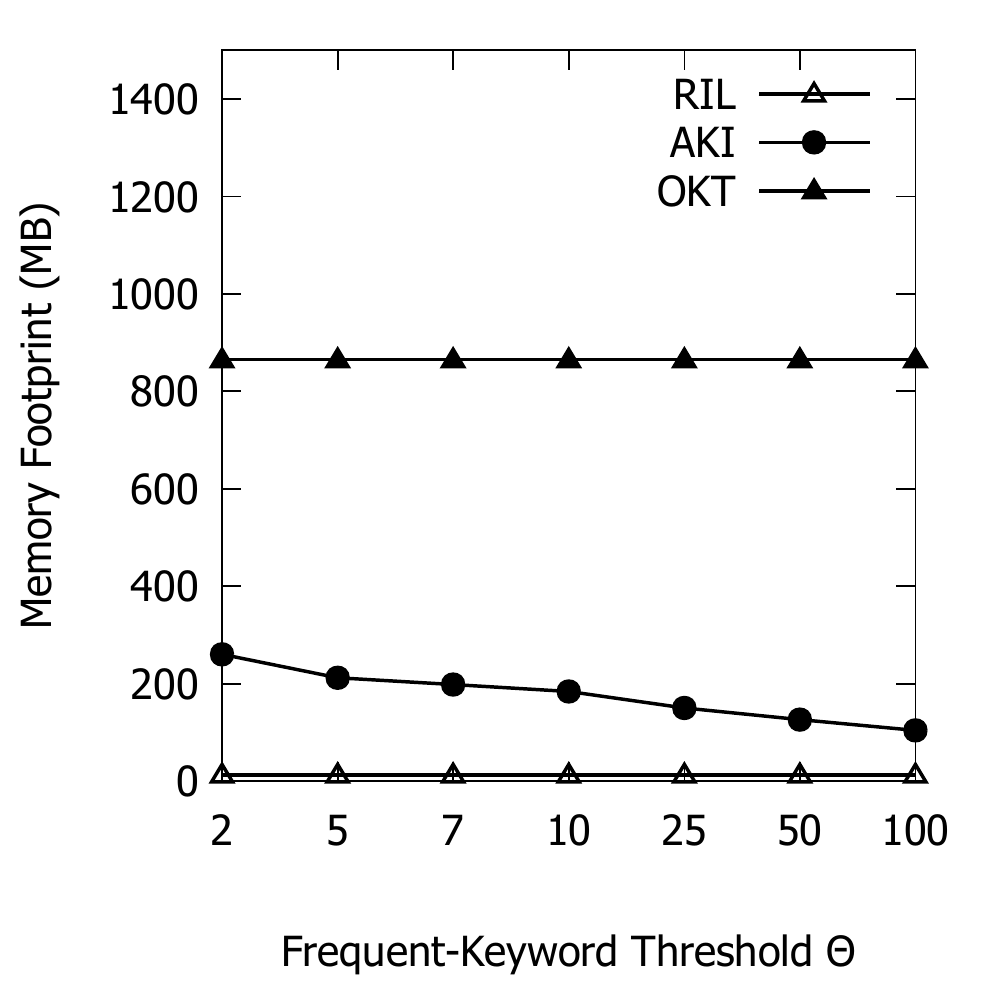}}
             \subfigure[Matching time FAST]{	\includegraphics[width=1.3in]	{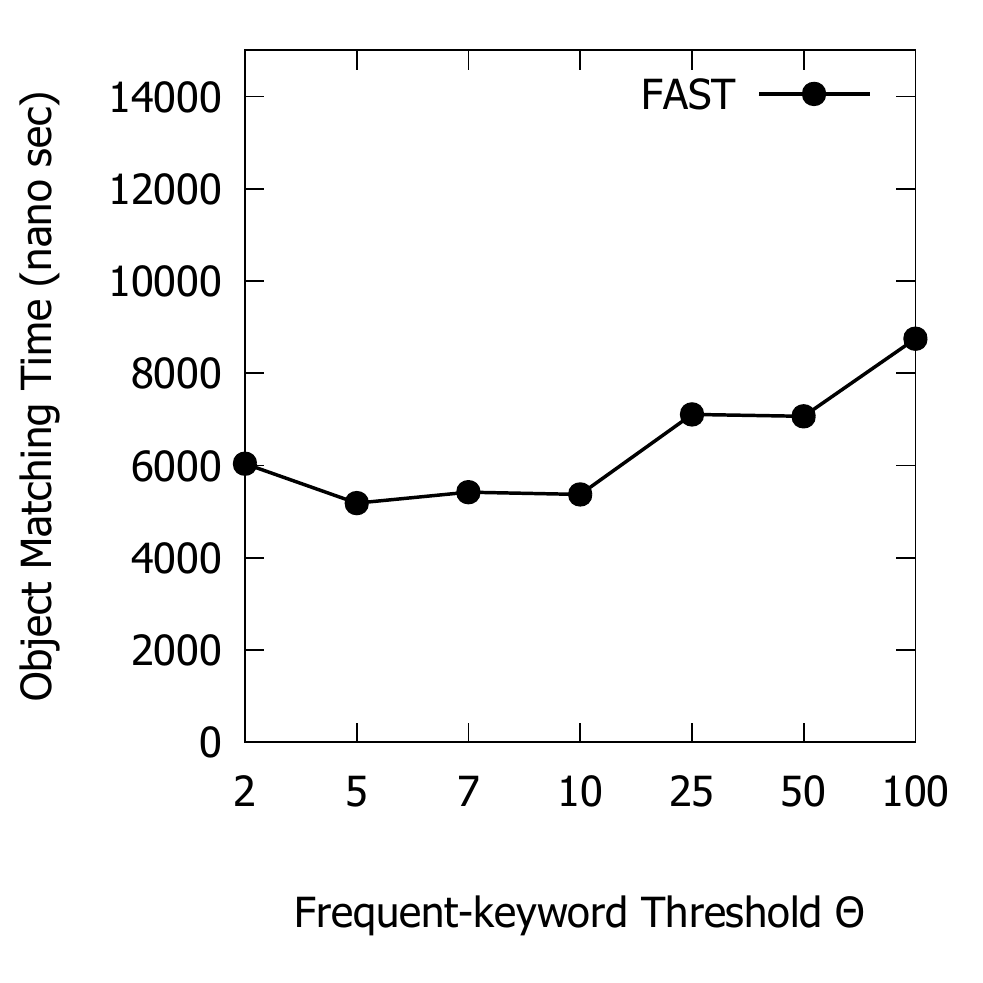}}
        \subfigure[Memory footprint FAST]{	\includegraphics[width=1.3in]	{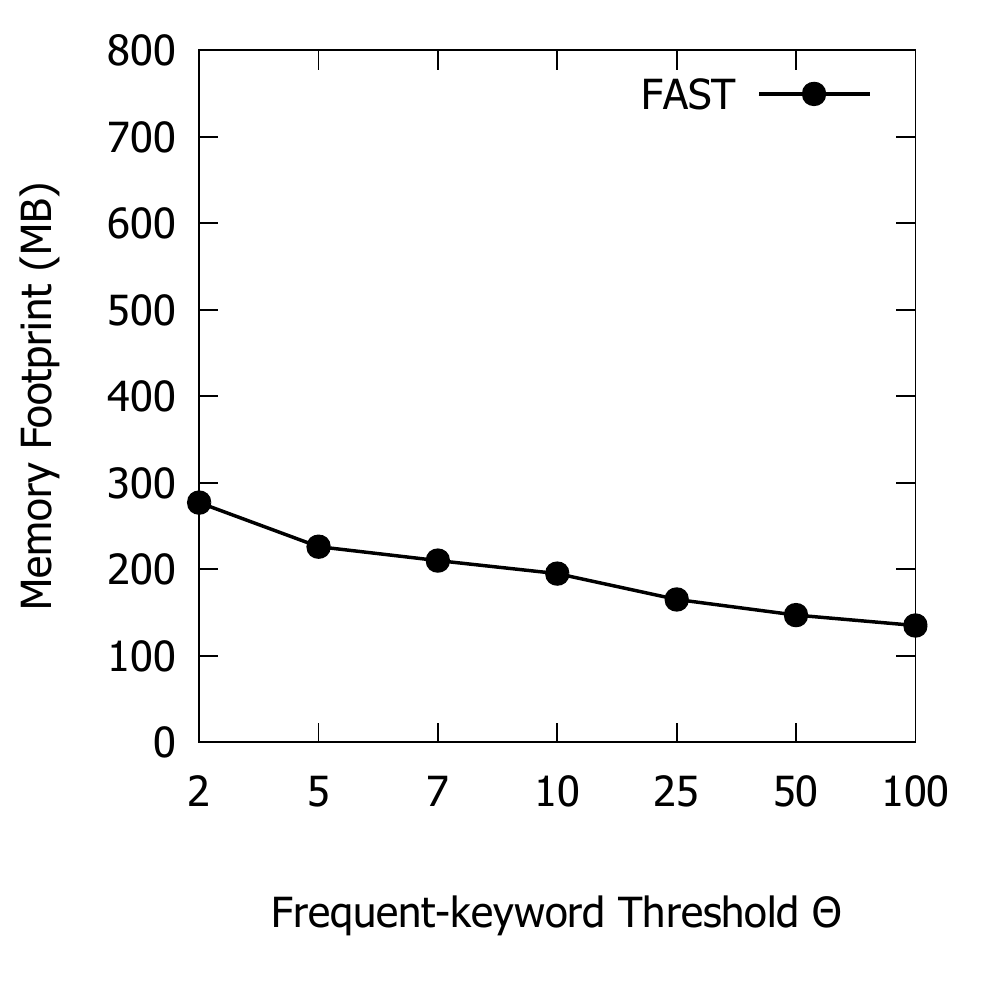}}
	  \caption{\textit{Frequent-keyword} threshold $\theta$ }\label{fig:expandablekeywordtheta}
      \end{minipage}
\begin{minipage}{.19\textwidth}
\raggedright
            	\subfigure[FAST Granularity]{\includegraphics[width=1.3in]	{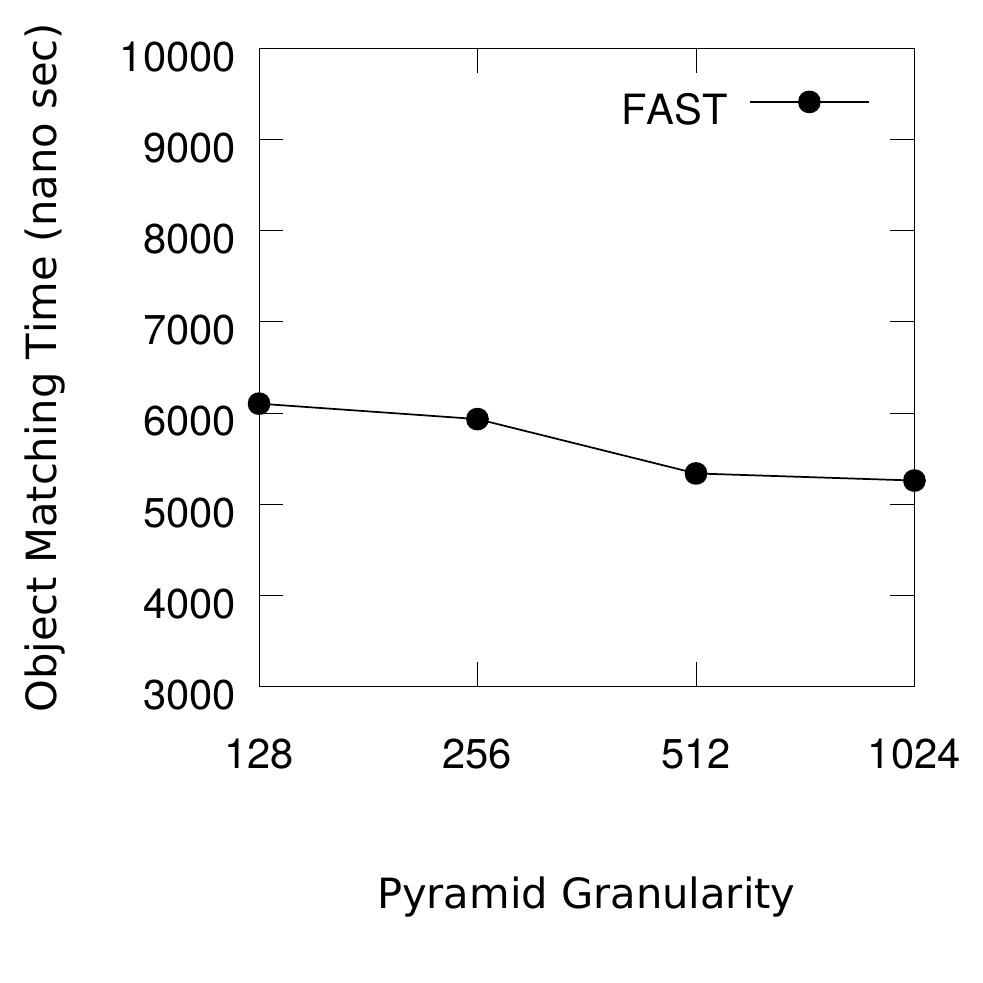}}
	  \caption{ Granularity}\label{fig:granularity}
      \end{minipage}
 \end{figure*}%
\textbf{Datasets.} Two real datasets, namely, \textit{Tweets} and \textit{Places}, and three synthetic datasets, namely, \textit{SpatialUni}, \textit{SpatialSkew}, and \textit{TextUni}, are used in the experimental evaluation. The Tweets dataset consists of 30 million geo-tagged tweets located inside the United States. These tweets are collected over the period from January 2014 to March 2015. The Places~\cite{pois} dataset contains 12.9 million public places inside the United States. Each entry 
in the Places dataset includes the geo-location and the set of keywords describing a specific place represented by the entry. Frequencies of keywords in both datasets follow a Zipfian distribution, as illustrated in Figure~\ref{fig:zipf}. Table~\ref{tab:dataset} summarizes the details of the real datasets used in the experiments.

The SpatialUni and the SpatialSkew synthetic datasets change the spatial location of entries in the Tweets dataset to follow a uniform and a skewed Gaussian distribution, respectively.  The TextUni dataset uses the spatial locations of entries in the Tweets dataset. However, keywords in the TextUni dataset are chosen uniformly from the vocabulary of the Tweets dataset, i.e., the frequencies of keywords follow a uniform distribution. We use the TextUni dataset to study the performance of FAST under a textual distribution that is not Zipfian to demonsrate that FAST is able to maintain its performance under various textual distributions.\\
\textbf{Query Workload}. Entries in datasets are used to construct spatial-keyword filter queries. The geo-location of a dataset entry is used as the center of the spatial range of a query. The default spatial range is a random value between 0\% and 1\% of the entire spatial range. The default number of query keywords is 3. 
Table~\ref{tab:experiments} summarizes
the query workload used in the experimental evaluation. \\
\textbf{Object Workload}. The AP-tree requires a training phase. We use 100K random dataset entries as historical training data. To measure the average matching time, we stream 100k data objects generated from the dataset entries against the indexed spatial-keyword queries. In the SpatialSkew dataset, we generate two synthetic object workloads, namely, SpatialSkewL and SpatialSkewO, where the spatial locations of objects in SpatialSkewL follow the same Guassian distribution as the one for the indexed queries. The spatial locations of objects in SpatialSkewO are skewed away from the spatial locations of the indexed queries. 

All implementations are in Java 8. All experiments are conducted on a 64-bit virtual machine running Ubuntu Linux 16.04. This virtual machine is allocated 16 cores each clocked at 2.6MHz. The total memory of the virtual machine is 49GB. 
The source code of the AP-tree has been provided by the authors of the AP-tree index. We set the parameters of the AP-tree according to the default values recommended by the authors of the AP-tree index~\cite{wang2015ap}. In our experiments, we report the object matching time, the query 
insertion time into the index, 
and the main-memory footprint of both FAST  and the AP-tree.

\subsection{Index Tuning}
In this section, we describe how to set the parameters of FAST. The main parameter in FAST is the \textit{frequent-keyword} threshold $\theta$. In this experiment, we study the performance of AKI and FAST under 
various
\textit{frequent-keyword} thresholds.\\
\noindent
\textbf{Performance of Textual Indexes.}
In Figure~\ref{fig:expandablekeywordtheta}, we study the effect of varying the \textit{frequent-keyword} threshold $\theta$ on the performance of AKI. We compare both the matching time and the memory footprint of 
AKI against 
both 
RIL and OKT. In Figure~\ref{fig:expandablekeywordtheta}(a), 
notice 
that AKI achieves  keyword matching time that 
is
comparable to 
that of 
OKT when $\theta \leq 10$ while having a memory footprint that is up to one third of that required by OKT. Notice that the performance of OKT and RIL is not affected by varying $\theta$ as both RIL and OKT do not have the \textit{frequent-keyword} threshold parameter. Increasing  $\theta$ increases the matching time and reduces the memory footprint of AKI. The reason is that as $\theta$ increases, the number of textual items attached to infrequent textual nodes increases. Matching textual items attached to infrequent textual nodes requires further validation. This validation increases the overall matching time. \\
\noindent
\textbf{Performance of FAST.}
In Figure~\ref{fig:expandablekeywordtheta}(c), 
observe 
that as we increase the \textit{frequent-keyword} threshold $\theta$, the matching time of FAST deteriorates. The reason is that the higher the \textit{frequent-keyword} threshold the 
longer 
the 
list
of queries attached to 
the
infrequent textual nodes, as in Figure~\ref{fig:expandablekeywordtheta}(a). This increases the textual validation time needed to verify the containment of query keywords within the keywords of 
the
streamed data objects. Figure~\ref{fig:expandablekeywordtheta}(d) 
demonstrates
that the smaller the \textit{frequent-keyword} threshold the higher the memory requirements of FAST. 

The reason is that having a small \textit{frequent-keyword} threshold results in marking more textual nodes as frequent, and demanding more memory for the splitting of their attached lists of queries. Figure~\ref{fig:expandablekeywordtheta} illustrates that using a \textit{frequent-keyword} threshold between 5 and 10 results in good matching time with moderate memory requirements in 
FAST. We set the \textit{frequent-keyword} threshold to 5 thourghout the rest of the experiments.
The formula of Equation~\ref{eq:theta} estimates that the worst case value of $\theta$ is $13.6$ that conforms with the simulation results in Figure~\ref{fig:expandablekeywordtheta}.\\
\noindent
\textbf{The Effect of Varying the Pyramid Granularity.}
Figure~\ref{fig:granularity}
illustrates
the matching time of FAST while varying the finest granularity of the spatial pyramid. 
From the figure, increasing the granularity of the pyramid within FAST improves the matching time initially. 
Then, increasing the granularity 
further
does not offer further improvement. 
Because of this observation, 
we set the granularity of FAST to 512 as increasing the finest granularity beyond 512 
does not improve 
the matching performance.
\begin{figure}
 		\centering
            \subfigure[Cleaning overhead]{	\includegraphics[width=1.3in]	{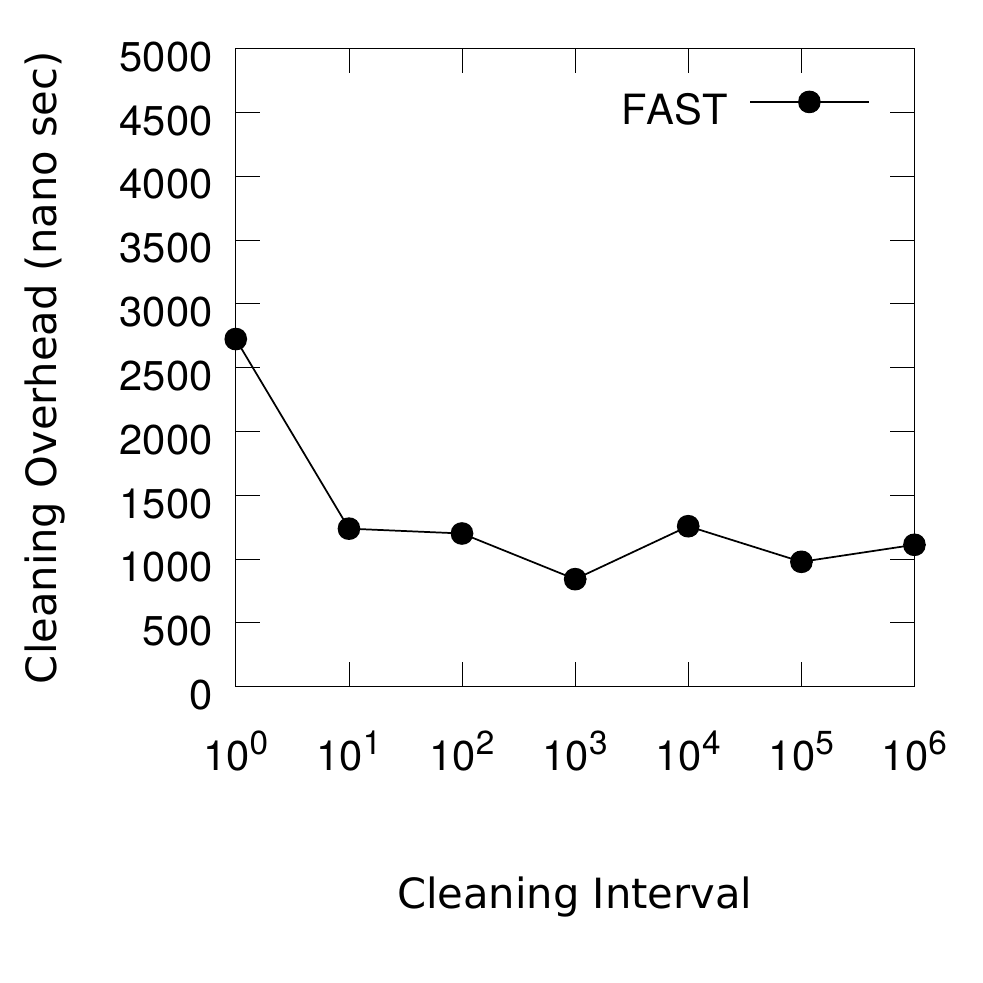}}
        \subfigure[Memory footprint]{	\includegraphics[width=1.3in]	{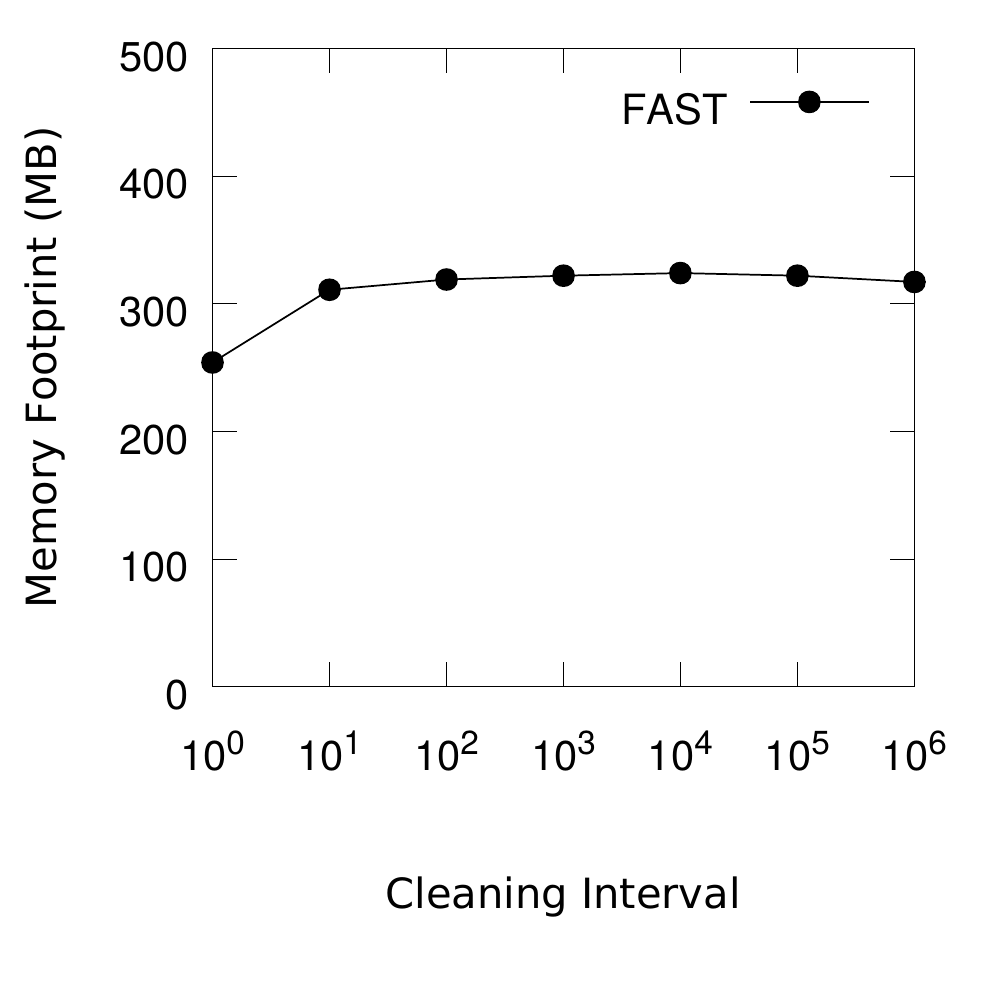}}
	  \caption{The effect of varying the cleaning interval $I$  }\label{fig:cleaning}
\end{figure}

\begin{figure*}[t!]
\centering
\begin{minipage}{.6\textwidth}
\raggedleft
        \subfigure[Matching time]{	\includegraphics[width=1.3in]{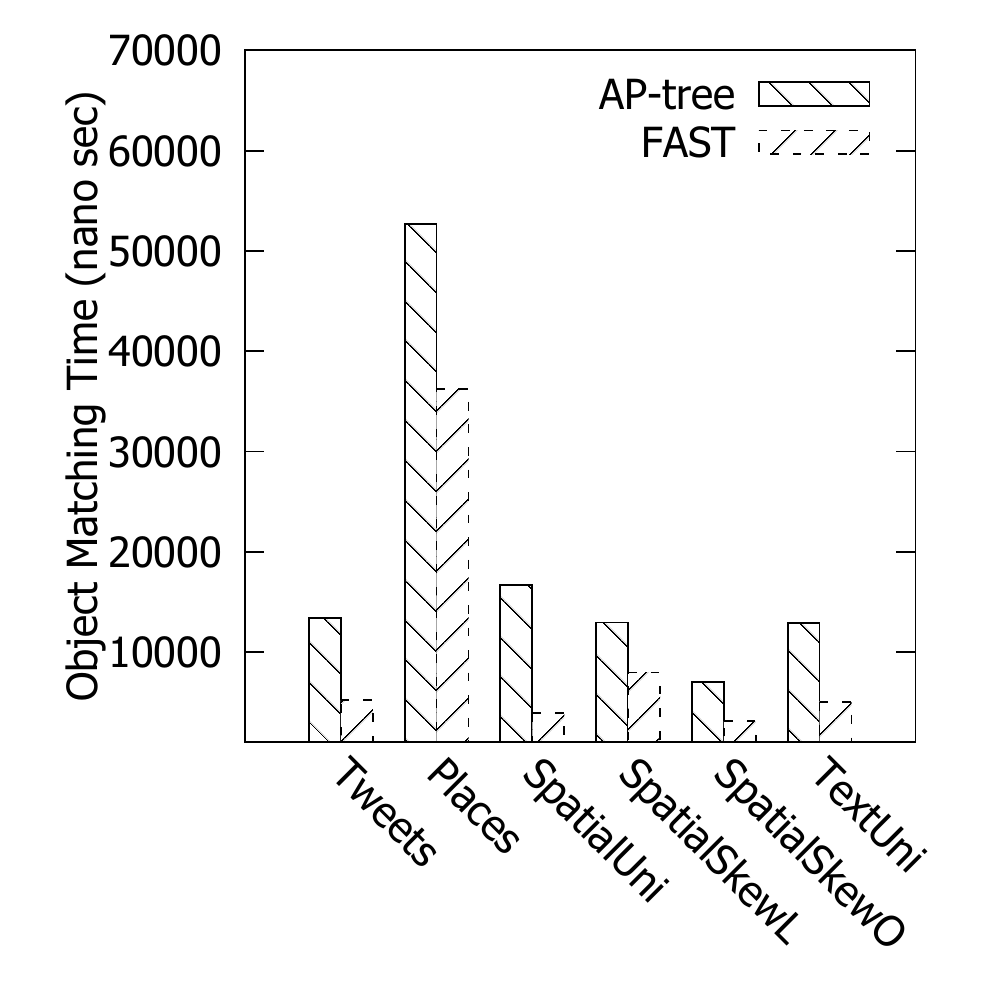}}
                 \subfigure[Indexing time]{	\includegraphics[width=1.3in]	{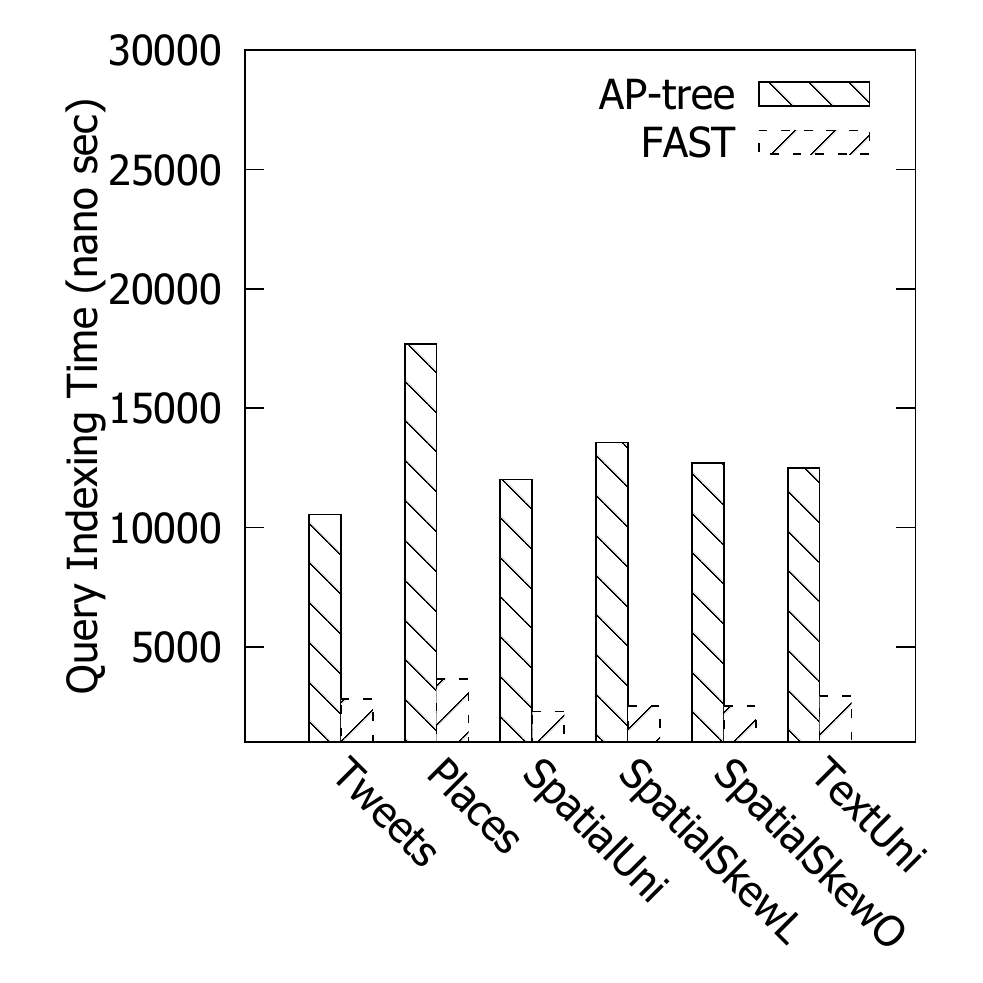}}
         \subfigure[Memory footprint ]{	\includegraphics[width=1.3in]	{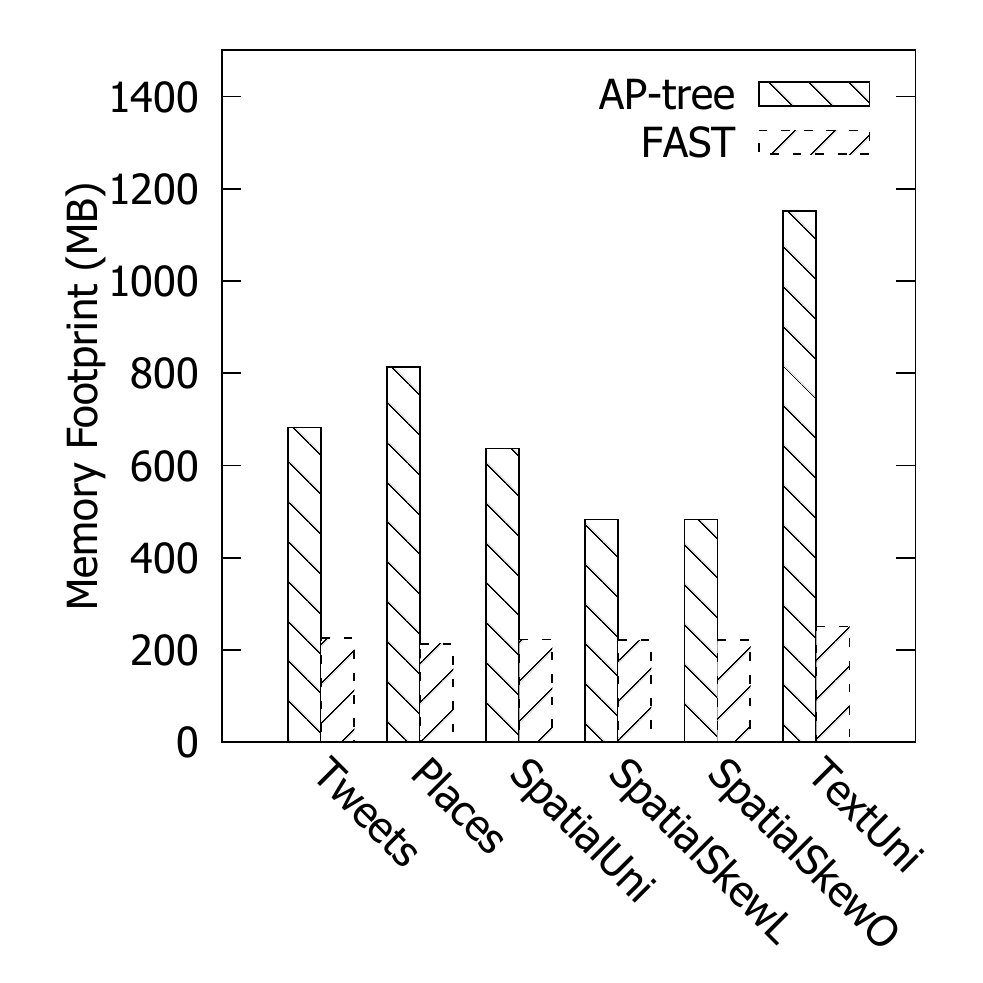}}
        \caption{Performance under different datasets}\label{fig:dataseteffect}
  \end{minipage}%
\begin{minipage}{.4\textwidth}
\raggedright
        \subfigure[Matching time]{	\includegraphics[width=1.3in]{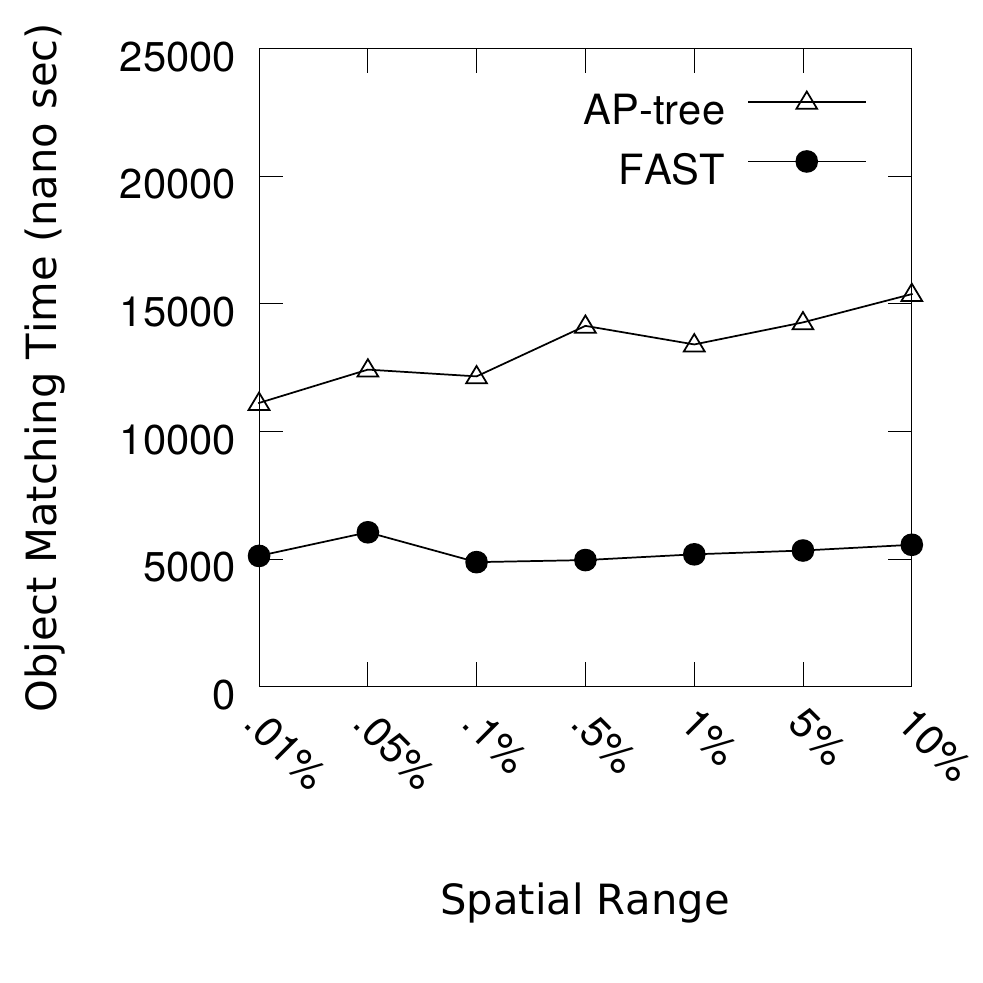}}
         \subfigure[Indexing time]{	\includegraphics[width=1.3in]	{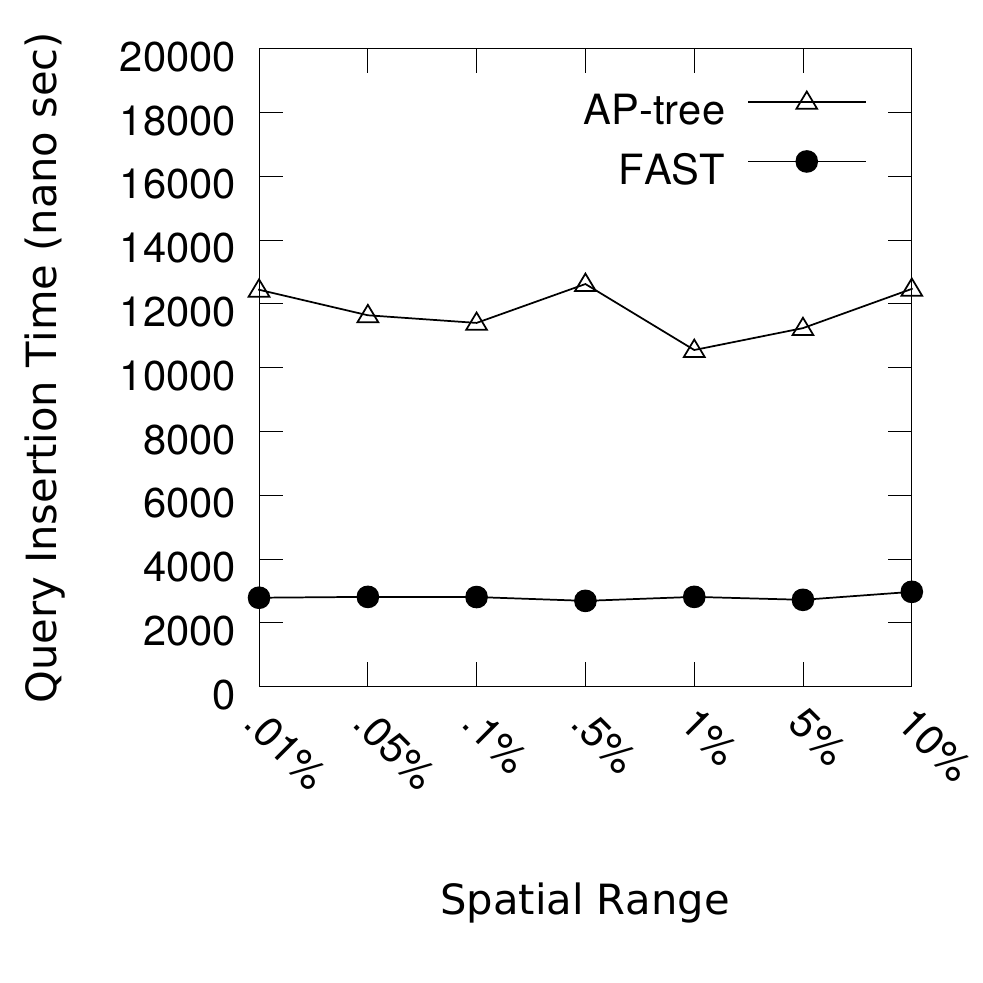}}
        \caption{The effect of the spatial range}\label{fig:spatialrangeeffect}
  \end{minipage}%
\end{figure*}
\begin{figure*}[t!]
\centering
\begin{minipage}{.4\textwidth}
\raggedleft
        \subfigure[Matching time]{	\includegraphics[width=1.3in]{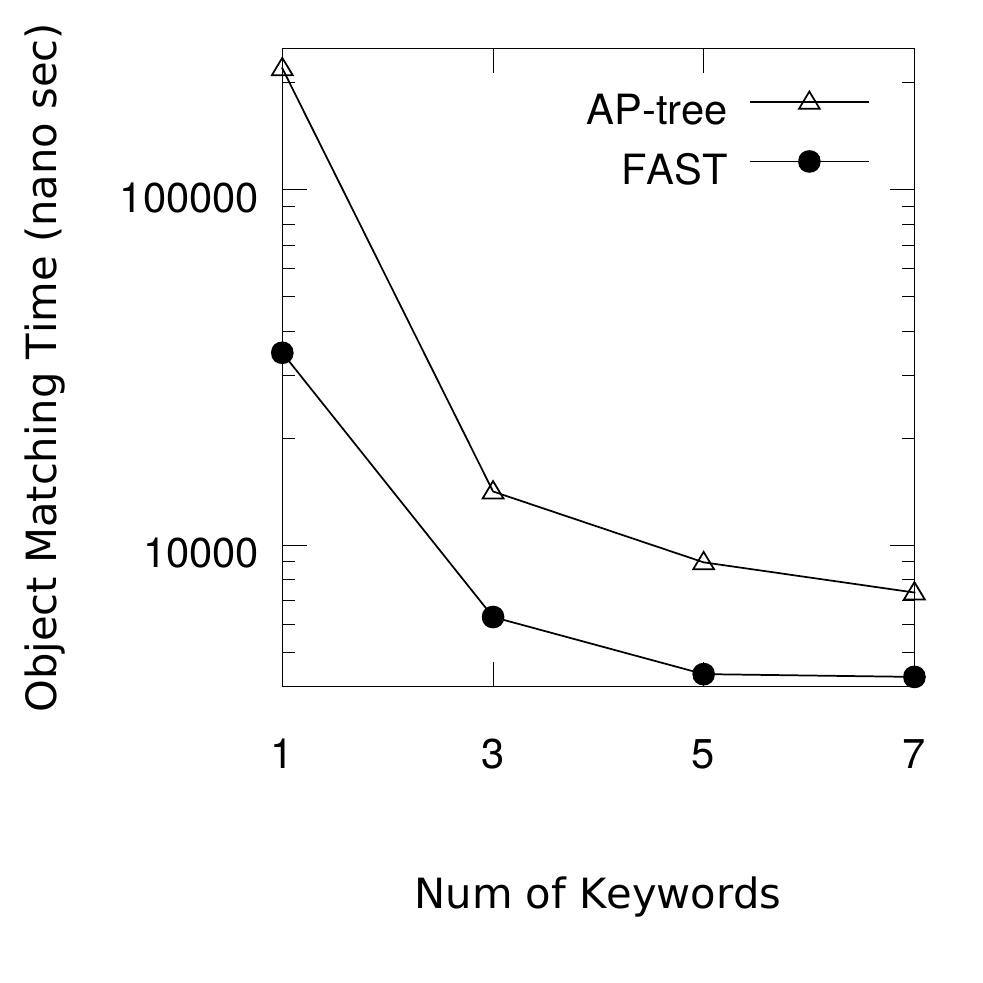}}
                 \subfigure[Indexing time]{	\includegraphics[width=1.3in]	{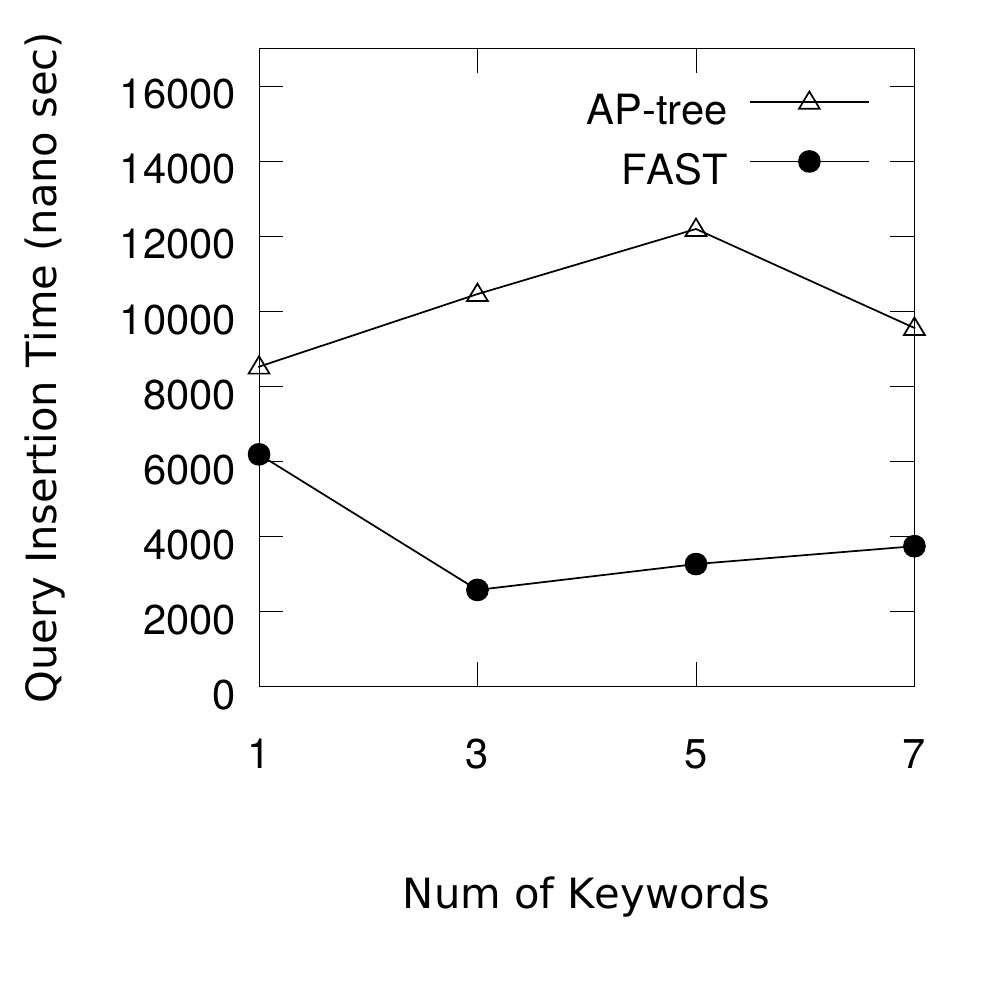}}
        \caption{The effect of the number of keywords}\label{fig:numebrofkeywords}
  \end{minipage}%
\begin{minipage}{.6\textwidth}
\raggedright
  \subfigure[Matching time]{		\includegraphics[width=1.3in]	{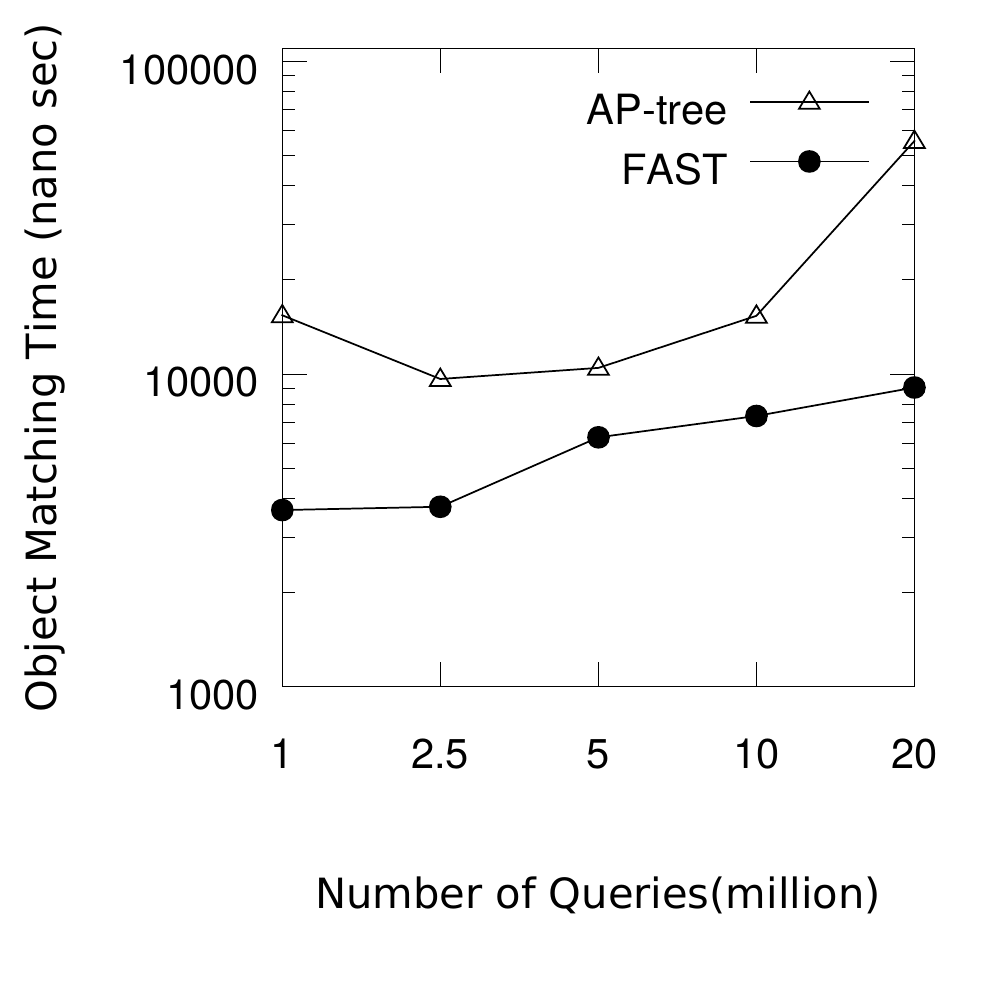}}
        \subfigure[Indexing time]{	\includegraphics[width=1.3in]{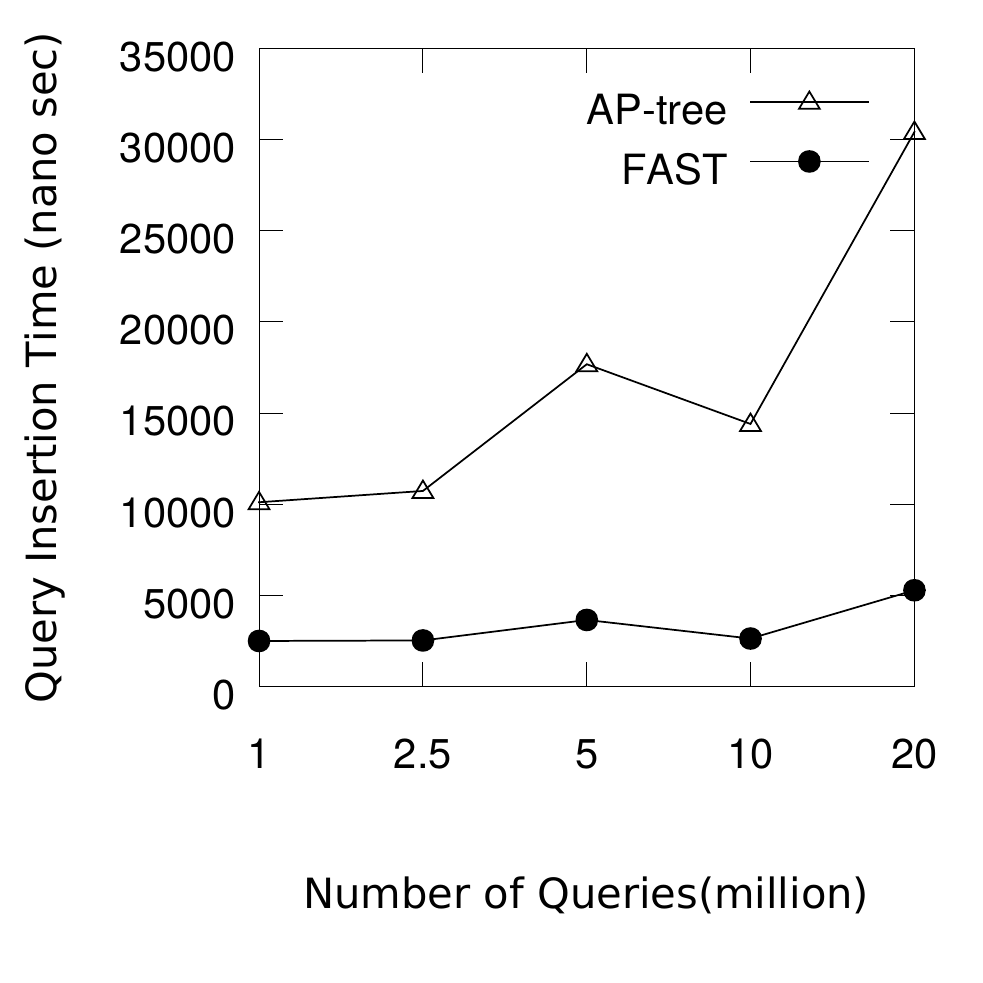}}
         \subfigure[Memory footprint ]{	\includegraphics[width=1.3in]	{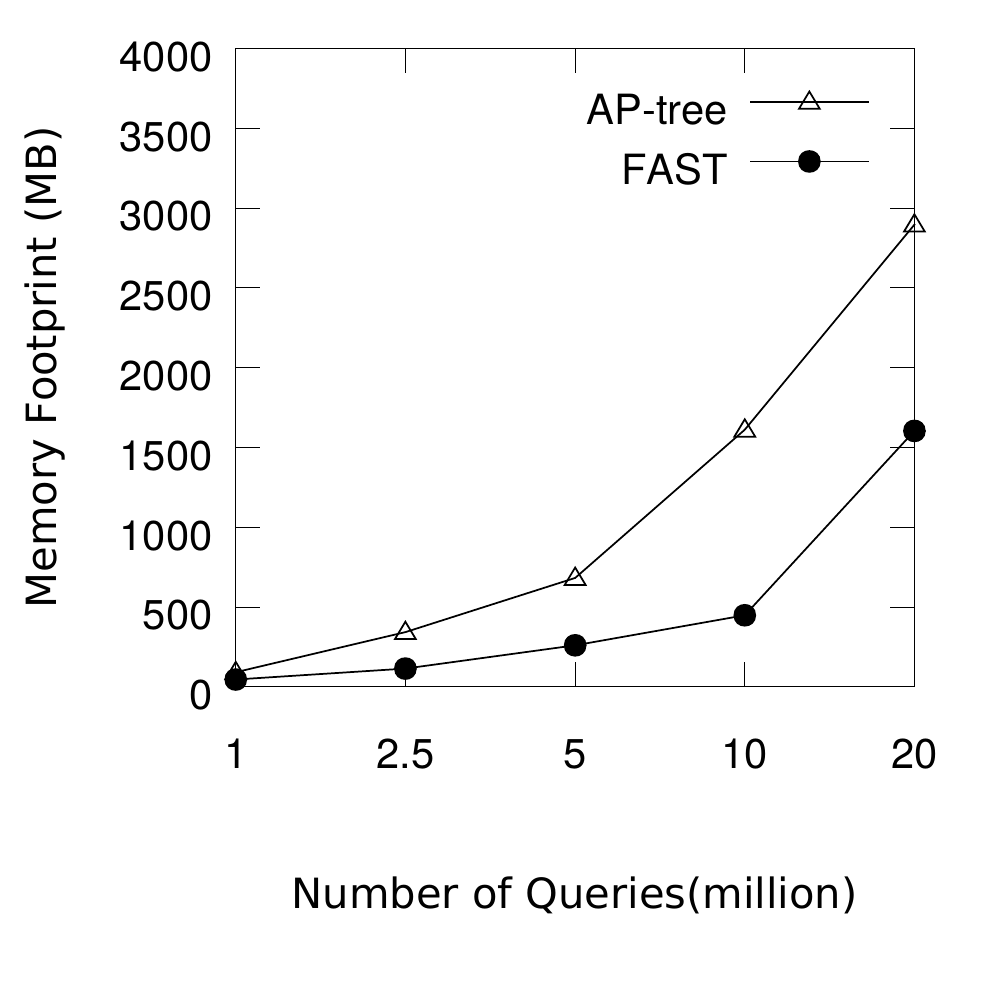}}
        \caption{Performance under varying number of indexed queries}\label{fig:scalability}
  \end{minipage}%
\end{figure*}
\noindent
\textbf{The Cleaning Overhead.}
In order to remove the expired queries in FAST, cells of FAST are visited periodically to be cleaned, i.e., every $I$ time units a cell is visited to be cleaned, as described in Section~\ref{subsec:cleaning}. Figure~\ref{fig:cleaning} illustrates the effect of varying the cleaning interval $I$ on the memory footprint of FAST and the average cleaning overhead, i.e., the average time 
spent in
cleaning. Figure~\ref{fig:cleaning}(a) 
illustrates 
that the cleaning overhead decreases as the cleaning interval increases. Having a very small cleaning interval results in redundant visits to cells that have been recently cleaned. Figure~\ref{fig:cleaning}(b) illustrates that the memory footprint of FAST increases as the cleaning interval $I$ increases. In our experiments, we set the cleaning interval to 1000 time units as it achieves 
balance between the cleaning overhead and the memory footprint of FAST. 

\subsection{Performance Evaluation}
In this section, we study the performance of FAST under various query workloads.\\
\noindent
\textbf{Performance using the Various Datasets.}
We compare the performance of FAST against that of the AP-tree using both the real and the synthetic datasets. Figure~\ref{fig:dataseteffect}(a) 
illustrates that FAST is up to 3x faster than the AP-tree in terms of object matching time. The reason is that FAST accounts for spatial and textual selectivities at the keyword level, as described in Section~\ref{sec:structure}. 
Figure~\ref{fig:dataseteffect}(b) illustrates that FAST is up to 5x faster than the AP-tree in terms of query indexing time. The reason is that FAST benefits from the \textit{frequent-keyword} threshold to account for the spatial and textual selectivities of data. However, the AP-tree uses an expensive cost formula to arbitrate between the spatial and textual indexing. In terms of the memory footprint,  Figure~\ref{fig:dataseteffect}(c) illustrates 
that FAST requires up to 3x less memory than that of the AP-tree. The reason is that FAST integrates AKI with spatial-cell sharing to reduce the size of the textual indexes, and to limit the replication of queries. However, the AP-tree is based on the memory intensive OKT and does not impose any restrictions on the replication of the indexed queries among the index cells. Notice that FAST maintains its performance advantages over the AP-tree under different synthetic distributions of the spatial and textual aspects of the data objects and queries.\\
\noindent
\textbf{The Effect of Varying the Spatial Range.}
In this experiment, we vary the spatial ranges of the queries from .01\% to 10\% of the entire spatial range. Figure~\ref{fig:spatialrangeeffect} illustrates the object matching time and the query indexing time for both the AP-tree and FAST. 
From the figure, observe
that FAST maintains its performance advantage against the AP-tree for both the object matching 
and the query indexing (insertion) times.\\
\noindent
\textbf{The Effect of Varying the Number of Keywords.}
In this experiment, we measure the \textit{object matching time} and \textit{the query indexing time} when changing the number of keywords in 
the 
indexed queries from 1 to 7. Figure~\ref{fig:numebrofkeywords} illustrates that FAST remains up to 3x faster than the AP-tree in terms of the object matching time and up to 5x faster in terms 
of the 
query indexing time.\\
\noindent
\textbf{The Scalability of FAST.}
In this experiment, we demonstrate the scalability of FAST against that of the AP-tree when increasing the number of indexed queries from 1 million to 20 million. Figure~\ref{fig:scalability} illustrates 
that
FAST maintains its performance advantage against the AP-tree. When increasing the number of indexed queries, FAST remains 3x faster than the AP-tree in object matching time, 5x faster than the AP-tree in the query indexing time, and requires one third of the main-memory required by the AP-tree.


\section{Related Work}\label{sec:relatedwork}
We classify the related work into the following categories: (1)~spatio-textual indexing, (2)~publish/subscribe systems, and (3)~superset containment search.

\noindent
{\bf Spatio-Textual Indexing.}
Recently, several spatio-textual indexes have been proposed to answer snap-shot queries over spatio-textual data. Examples of these queries include the filter, top-k, and collective group queries. Chen et al.~\cite{chen2013spatial} surveys spatio-textual indexes and benchmarks their performance under various spatio-textual queries.  
The most relevant indexes are the IQ-tree~\cite{chen2006efficient} and the R$^t$-tree~\cite{li2013location}. These indexes are mainly disk-based and have been outperformed by the AP-tree~\cite{wang2015ap}.

\noindent
{\bf Publish/Subscribe Systems.}
One main use case of FAST is 
in
location-aware publish/subscribe systems. Publish/subscribe systems maintain subscriptions for long durations and match incoming messages against stored subscriptions. Publish/subscribe systems can be categorized according to their matching approach into the following categories: (1)~content-based~\cite{zhang2014efficient}, (2)~TopK-similarity-based~\cite{shraer2013top}, and (3)~location-aware~\cite{bao2012geofeed}. These publish/subscribe systems do not simultaneously account for the spatial and textual properties of subscriptions and messages.
Recently, several spatio-textual publish/subscribe systems~\cite{wang2015ap,hu2015location} have been proposed. To the best of our knowledge, the AP-tree~\cite{wang2015ap} is the most relevant work for indexing continuous spatio-textual queries in a streaming environment. 


\noindent
{\bf Superset Containment Search.}
AKI addresses the problem of superset containment search, where it is required to retrieve indexed items with keywords that are fully contained in the search keywords. Several indexes have been proposed to address the superset containment problem, e.g.,~\cite{zobel2006inverted,yan1994index,terrovitis2011efficient,terrovitis2006combination}. OKT~\cite{yan1994index} and RIL~\cite{zobel2006inverted} are the most adopted structures for superset containment search~\cite{hmedeh2012subscription}. Terrovitis et al.~\cite{terrovitis2011efficient,terrovitis2006combination} present two structures for superset containment search. However, these structures are mainly disk-based and require knowing the frequencies of the entire vocabulary. AKI is a main-memory index and does not assume prior knowledge of the frequencies of keywords.
\section{Conclusion}\label{sec:conculsion}
In this paper, we introduce FAST; a Frequency-Aware Spatio-Textual access method for indexing continuous spatio-textual filter queries in a streaming environment. FAST automatically accounts for both the spatial and textual selectivities of indexed queries to improve the indexing and searching performance. FAST integrates the spatial pyramid with a new textual index, and supports a cell-sharing technique that reduces the memory required by the index. FAST uses a light-weight 
lazy-cleaning
mechanism to remove 
the
expired queries and to reflect changes in the frequencies of the keywords of 
the
indexed queries.  Extensive experimental evaluation using real and synthetic datasets 
demonstrates
that FAST is up to 3x faster in search time and 5x faster in indexing time than the state-of-the-art index. Also, FAST requires up to 3x less memory than the state-of-the-art index. 

\begin{small}
{
\footnotesize{
\bibliographystyle{IEEEtran}
\bibliography{IEEEabrv,IEEEexample}}

\begin{thebibliography}{10}
\providecommand{\url}[1]{#1}
\csname url@samestyle\endcsname
\providecommand{\newblock}{\relax}
\providecommand{\bibinfo}[2]{#2}
\providecommand{\BIBentrySTDinterwordspacing}{\spaceskip=0pt\relax}
\providecommand{\BIBentryALTinterwordstretchfactor}{4}
\providecommand{\BIBentryALTinterwordspacing}{\spaceskip=\fontdimen2\font plus
\BIBentryALTinterwordstretchfactor\fontdimen3\font minus
  \fontdimen4\font\relax}
\providecommand{\BIBforeignlanguage}[2]{{%
\expandafter\ifx\csname l@#1\endcsname\relax
\typeout{** WARNING: IEEEtran.bst: No hyphenation pattern has been}%
\typeout{** loaded for the language `#1'. Using the pattern for}%
\typeout{** the default language instead.}%
\else
\language=\csname l@#1\endcsname
\fi
#2}}
\providecommand{\BIBdecl}{\relax}
\BIBdecl

\bibitem{wang2015ap}
X.~Wang, Y.~Zhang, W.~Zhang, X.~Lin, and W.~Wang, ``Ap-tree: Efficiently
  support continuous spatial-keyword queries over stream,'' in \emph{ICDE},
  2015, pp. 1107--1118.

\bibitem{geotaggedtweets}
``Geotagged tweets,'' \url{http://www.futurity.org/tweets-give-info-location/},
  2017.

\bibitem{foursquare}
``Foursquare,'' \url{https://foursquare.com/about}, 2017.

\bibitem{mahmood2015tornado}
A.~R. Mahmood, A.~M. Aly, T.~Qadah, E.~K. Rezig, A.~Daghistani, A.~Madkour,
  A.~S. Abdelhamid, M.~S. Hassan, W.~G. Aref, and S.~Basalamah, ``Tornado: A
  distributed spatio-textual stream processing system,'' \emph{PVLDB}, vol.~8,
  no.~12, pp. 2020--2023, 2015.

\bibitem{yan1994index}
T.~W. Yan and H.~Garc{\'\i}a-Molina, ``Index structures for selective
  dissemination of information under the boolean model,'' \emph{TODS}, vol.~19,
  no.~2, pp. 332--364, 1994.

\bibitem{konig2009data}
A.~C. K{\"o}nig, K.~Church, and M.~Markov, ``A data structure for sponsored
  search,'' in \emph{ICDE}, 2009, pp. 90--101.

\bibitem{chen2013efficient}
L.~Chen, G.~Cong, and X.~Cao, ``An efficient query indexing mechanism for
  filtering geo-textual data,'' in \emph{SIGMOD}, 2013, pp. 749--760.

\bibitem{li2013location}
G.~Li, Y.~Wang, T.~Wang, and J.~Feng, ``Location-aware publish/subscribe,'' in
  \emph{SIGKDD}, 2013, pp. 802--810.

\bibitem{guttman1984r}
A.~Guttman, \emph{R-trees: a dynamic index structure for spatial
  searching}.\hskip 1em plus 0.5em minus 0.4em\relax ACM, 1984, vol.~14, no.~2.

\bibitem{finkel1974quad}
R.~A. Finkel and J.~L. Bentley, ``Quad trees a data structure for retrieval on
  composite keys,'' \emph{Acta informatica}, vol.~4, no.~1, pp. 1--9, 1974.

\bibitem{zobel2006inverted}
J.~Zobel and A.~Moffat, ``Inverted files for text search engines,'' \emph{ACM
  computing surveys}, vol.~38, no.~2, p.~6, 2006.

\bibitem{hmedeh2012subscription}
Z.~Hmedeh, H.~Kourdounakis, V.~Christophides, C.~Du~Mouza, M.~Scholl, and
  N.~Travers, ``Subscription indexes for web syndication systems,'' in
  \emph{EDBT}.\hskip 1em plus 0.5em minus 0.4em\relax ACM, 2012, pp. 312--323.

\bibitem{powers1998applications}
D.~M. Powers, ``Applications and explanations of zipf's law,'' in
  \emph{Proceedings of the joint conferences on new methods in language
  processing and computational natural language learning}, 1998, pp. 151--160.

\bibitem{knuth1968art}
D.~E. Knuth, \emph{The art of computer programming}.\hskip 1em plus 0.5em minus
  0.4em\relax Addison-Wesley, 1968, vol.~3.

\bibitem{aref1990efficient}
W.~G. Aref and H.~Samet, ``Efficient processing of window queries in the
  pyramid data structure,'' in \emph{PODS}.\hskip 1em plus 0.5em minus
  0.4em\relax ACM, 1990, pp. 265--272.

\bibitem{magdy2015towards}
A.~Magdy and M.~F. Mokbel, ``Towards a microblogs data management system,'' in
  \emph{MDM}, vol.~1, 2015, pp. 271--278.

\bibitem{lee2015processing}
T.~Lee, J.-w. Park, S.~Lee, S.-W. Hwang, S.~Elnikety, and Y.~He, ``Processing
  and optimizing main memory spatial-keyword queries,'' \emph{PVLDB}, vol.~9,
  no.~3, pp. 132--143, 2015.

\bibitem{enderton2001mathematical}
H.~Enderton and H.~B. Enderton, \emph{A mathematical introduction to
  logic}.\hskip 1em plus 0.5em minus 0.4em\relax Academic press, 2001.

\bibitem{pois}
``{Places dataset},''
  \url{https://archive.org/details/2011-08-SimpleGeo-CC0-Public-Spaces}, 2017.

\bibitem{chen2013spatial}
L.~Chen, G.~Cong, C.~S. Jensen, and D.~Wu, ``Spatial keyword query processing:
  An experimental evaluation,'' in \emph{VLDB}, vol.~6, no.~3, 2013, pp.
  217--228.

\bibitem{chen2006efficient}
Y.-Y. Chen, T.~Suel, and A.~Markowetz, ``Efficient query processing in
  geographic web search engines,'' in \emph{SIGMOD}.\hskip 1em plus 0.5em minus
  0.4em\relax ACM, 2006, pp. 277--288.

\bibitem{zhang2014efficient}
D.~Zhang, C.-Y. Chan, and K.-L. Tan, ``An efficient publish/subscribe index for
  e-commerce databases,'' \emph{PVLDB}, vol.~7, no.~8, pp. 613--624, 2014.

\bibitem{shraer2013top}
A.~Shraer, M.~Gurevich, M.~Fontoura, and V.~Josifovski, ``Top-k
  publish-subscribe for social annotation of news,'' \emph{PVLDB}, vol.~6,
  no.~6, pp. 385--396, 2013.

\bibitem{bao2012geofeed}
J.~Bao, M.~F. Mokbel, and C.-Y. Chow, ``Geofeed: A location aware news feed
  system,'' in \emph{ICDE}.\hskip 1em plus 0.5em minus 0.4em\relax IEEE, 2012,
  pp. 54--65.

\bibitem{hu2015location}
H.~Hu, Y.~Liu, G.~Li, J.~Feng, and K.-L. Tan, ``A location-aware
  publish/subscribe framework for parameterized spatio-textual subscriptions,''
  in \emph{ICDE}.\hskip 1em plus 0.5em minus 0.4em\relax IEEE, 2015, pp.
  711--722.

\bibitem{terrovitis2011efficient}
M.~Terrovitis, P.~Bouros, P.~Vassiliadis, T.~Sellis, and N.~Mamoulis,
  ``Efficient answering of set containment queries for skewed item
  distributions,'' in \emph{EDBT}, 2011, pp. 225--236.

\bibitem{terrovitis2006combination}
M.~Terrovitis, S.~Passas, P.~Vassiliadis, and T.~Sellis, ``A combination of
  trie-trees and inverted files for the indexing of set-valued attributes,'' in
  \emph{CIKM}, 2006, pp. 728--737.

\end{thebibliography}
}
\end{small}

\appendix

\begin{figure}
	\centering	\includegraphics[width=1.3in]{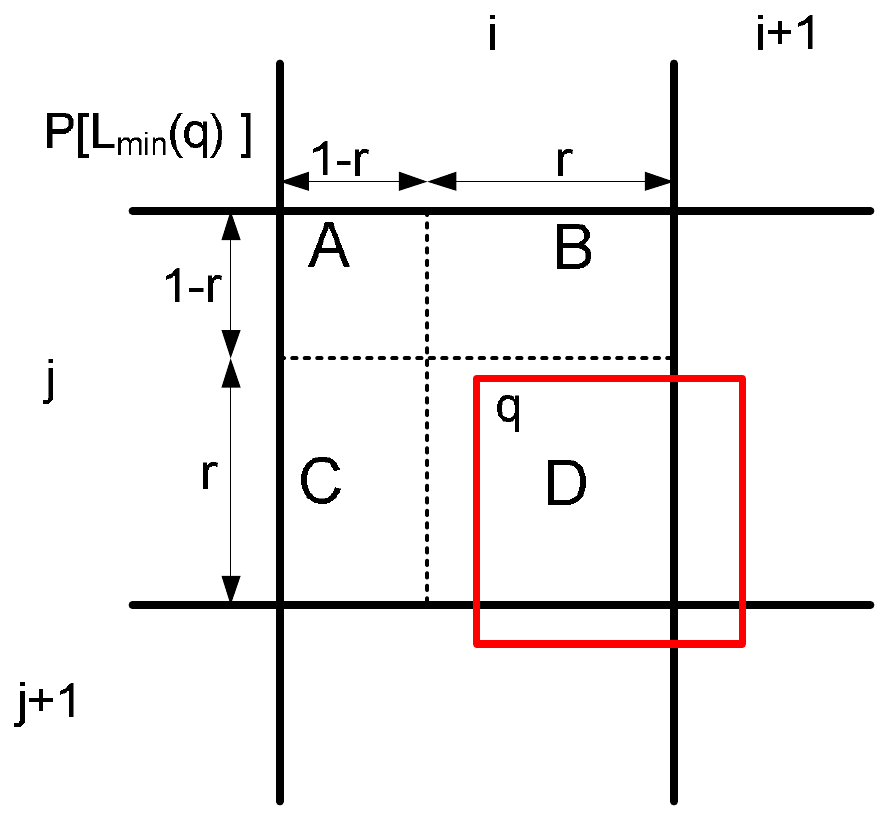}	\caption{Expected query replication.}	\label{fig:averagequeryreplication}
\end{figure}
\section{Expected Query Replication}\label{app:replication}

In this section, we estimate the expected replication of queries when indexed at their lowest allowed pyramid levels, i.e., $E_{rep}(L_{min}(q))$. As described in Section~\ref{sec:structure}, a query can descend down to  Level 
$L_{min}(q)$, where 
$SideLen(L_{min}(q))$ is strictly greater than the side length of Query $q$, i.e., $q.r$.  The side length of a query is calculated using  Equation~\ref{eq:sidelength}, and the side length of pyramid nodes at any given level is calculated using Equation~\ref{eq:step}. 

To simplify the analysis, we assume that 
a
pyramid node 
has
a unit side length, i.e., $SideLen(L_{min}(q))$= 1. Notice that $SideLen(L_{min}(q)) \geq q.r>  SideLen(L_{min}(q))/2$, i.e., $1 \geq q.r> .5$ for cells with unit side length. To find the expected replication, we assume that the side range $q.r$ is a random value in the range $].5,1]$. Figure~\ref{fig:averagequeryreplication} 
gives
the number of replications of Query $q$ in pyramid nodes at level $L_{min}(q)$. For Pyramid $P$, the replication of Query q can be determined by the placement of the top-left corner of $q$ in Cell $P[L_{min}(q)][i][j]$, where $i,\ j$ are the coordinates of the point $(q.x_{min},q.y_{max})$. The spatial range of Cell $P[L_{min}][i][j]$ can be divided into the regions: $A,\ B,\ C$, and $D$. 
he replication of $q$ in regions $A,\ B,\ C$, and $D$ depends on  the placement of $(q.x_{min},q.y_{max})$ across the regions of Cell $P[L_{min}(q)][i][j]$ is listed in Table~\ref{tab:replication}.

\begin{table}[!t]
\centering
    \caption{Replication of Query $q$.}     
    \label{tab:replication}
	{\renewcommand{\arraystretch}{1}%
    \begin{tabular}{|c|c|c|}
    \hline
       {\bfseries  Region } &{\bfseries Pr(region)}&{\bfseries Replication }\\
       \thickhline
       $A $ &$(1-r)^2 $ & 1\\
       \hline
       $B $ &$r(1-r) $ & 2\\
       \hline
       $C $ &$r(1-r) $ & 2\\
       \hline
       $D $ &$r^2 $ &4\\
       \hline
    \end{tabular}}
\end{table}
To calculate the expected replication, we integrate the expected replication of the queries across the regions
$A,\ B,\ C$, and $D$ as follows:\\
\noindent
$E_{rep}(L_{min}(q))=\dfrac{1}{1-.5}\int_{.5}^{1}$ $\sum replication\times Pr(region)$\\ 
$=\dfrac{1}{1-.5}\int_{.5}^{1} 4\times r^2+2\times 2\times r\times (1-r)+3\times 0+ 1\times (1-r)^2\ \  dr$
$=2\int_{.5}^{1} (1+r)^2\ \  dr$
=3.08 that is less than the worst case replication of $4$.\\
This analysis can be extended to queries indexed at a higher pyramid level ($L_{min}(q)+i$) as follows:\\
$E_{rep}(L_{min}(q)+i)=\dfrac{2}{2^{2i}}\int_{.5}^{1} (2^i+r)^2\ \  dr$\\
Notice that the query replication at levels higher than $L_{min}(q)$ is less than 3.08. For example, the query replication at pyramid level $L_{min}(q)+2$ is equal to 1.4 and at the top pyramid level is equal to 1. Furthermore, if indexed queries have side lengthes that follow a uniform distribution, where all possible query replications are equally likely to occure in a spatial pyramid with $n$ levels, the overall expected replication can estimated to be :\\
$E_{rep}=\dfrac{1}{n}\sum_{i=0}^{n-1}  \dfrac{2}{2^{2i}}\int_{.5}^{1} (2^i+r)^2\ \  dr$\\ that is equal to 1.27 when the number of levels $n$ is 9. The average query replication measured experimentally in FAST is 1.08, that is very close to the estimated query replication.

From this equation, $E_{rep}$ is equal to 1.27 when the number of levels $n$ is 9. The average query replication measured experimentally is 1.08 that is very close to the estimated query replication.

\end{document}